\documentclass[11pt]{article}
\usepackage{jheppub}
\usepackage{import}
\usepackage{TemplateMB}
\usepackage{physics}

\usepackage{microtype}

\usepackage{amsmath,amssymb,latexsym}
\usepackage{svg}
\usepackage{graphicx}
\usepackage{floatflt}
\let\originalleft\left
\let\originalright\right
\renewcommand{\left}{\mathopen{}\mathclose\bgroup\originalleft}
\renewcommand{\right}{\aftergroup\egroup\originalright}

\usepackage{mathtools}

\renewcommand{\tilde}{\widetilde}
\renewcommand{\bar}{\overline}

\def\cE{{\cal E}}

\def\cL{{\cal L}}
\def\cN{{\cal N}}
\def\cO{{\cal O}}
\def\cS{{\cal S}}
\def\cU{{\cal U}}
\def\cV{{\cal V}}

\let\obinom\binom
\renewcommand{\binom}[2]{{\textstyle\obinom{#1}{#2}}}

\def\dow{\partial}
\def\half{\frac12}
\def\SK{\text{SK}}

\def\lB{\left[}
\def\rB{\right]}

\def \de {\delta}

\def \om {\omega}

\def \ep {\epsilon}

\def \Ec {\mathcal{E}}

\def \Oc {\mathcal{O}}

\def \Sc {\mathcal{S}}

\def \Jc {\mathcal{J}}

\def \pr {\partial}
\def \ra {\rightarrow}

\def \beq { \begin{equation}}
\def \eeq {\end{equation}}

\DeclareMathOperator*{\Sch}{Sch}

\DeclareMathOperator{\SL}{SL}

\renewcommand\Re{\operatorname{Re}}
\renewcommand\Im{\operatorname{Im}}

\def \l {\left}
\def \r {\right}

\def \bra {\langle}
\def \ket {\rangle}

\usepackage{ulem}

\def\le{\left}
\def\ri{\right}
\newcommand\ov{\over}
\newcommand\p{\ensuremath{\partial}}
\newcommand{\es}[2] {\begin{equation} \ifx#1\empty\else\label{#1}\fi \begin{split} #2 \end{split} \end{equation}}

\def \coup {g} 

\def\<{\langle}
\def\>{\rangle}
\newcommand\sig{\sigma}
\newcommand\lam{\lambda}

\author[a]{Marta Bucca,}
\author[a]{Akash Jain,}
\author[a]{M\'ark Mezei,}    
\author[b,c]{Alexey Milekhin}                    
 
 \affiliation[a]{Mathematical Institute, University of Oxford, Woodstock Road, Oxford, OX2 6GG, United Kingdom}  

 \affiliation[b]{Institute for Quantum Information and Matter, California Institute of Technology, 1200 E. California Blvd., Pasadena, CA 91125, USA}  

\affiliation[c]{Department of Physics and Astronomy, University of Kentucky, 506 Library Drive, Lexington, KY 40506}  

\emailAdd{\{marta.bucca,akash.jain,mark.mezei\}@maths.ox.ac.uk}
\emailAdd{milekhin@uky.edu}

\begin{document}
\title{All-order fluctuating hydrodynamics of the SYK lattice}
\author{}
\date{}
\abstract{The SYK model has played an important role in recent developments in many-body quantum chaos. We study a spatially local generalisation of it: the SYK lattice. Starting from the nonlinear action of pseudo-Goldstone bosons that dominate its dynamics at low temperatures, in the long wavelength limit we reorganise this action as the effective field theory for fluctuating hydrodynamics, thereby showing how the hydrodynamic degrees of freedom embed into the microscopic description of the model. We compute the hydrodynamic effective action to high orders in the derivative expansion and determine all the corresponding transport coefficients. Hence this work derives hydrodynamics from the microscopic description of a strongly coupled quantum many-body system.
}

\maketitle

\section{Introduction and summary of results}

Quantum many-body systems exhibit rich and exotic phenomena, especially out-of-equilibrium. Just in the past few years, a plethora of novel phenomena have been discovered, such as measurement-induced phase transitions~\cite{Li:2018mcv,Skinner:2018tjl,Li:2019zju,Szyniszewski:2019pfo, Minato:2021hew, Muller:2021xxc, Buchhold:2022vyf,Dhar_2016,Turkeshi:2021mcz, Tang_2020,  Szyniszewski:2020fwl,Jian:2021hve, Nahum:2020hec, Botzung:2021pfn, Alberton:2020lfq, Altland:2021bqa}, scar states~\cite{Bernien_2017,turner2018weak,Su_2023,serbyn2021quantum,Moudgalya:2021xlu,Papic2022,Chandran:2022jtd,Pakrouski:2020hym,Milekhin:2023was}, strong-to-weak symmetry breaking~\cite{Buca:2012zz,Gu:2024wgc,Feng:2025ngy,Liu:2024mme}, and deep thermalisation~\cite{cotler23,choi23,ho22,ho23,chan2024,chang24,federica,Milekhin:2024imt}.
Typically, the late-time behaviour of physical systems is governed by hydrodynamics.\footnote{At least this is true in the presence of conservation laws and for simple observables, such as time-ordered correlators of local operators, which have been studied for the past two centuries. Very recently it was proposed that more complicated observables, such as operator size, can also obey hydrodynamic equations even without underlying conservation laws~\cite{vonKeyserlingk:2017dyr}.} 
One such example is the diffusion of energy/temperature we observe every day. However, diffusion and other conventional hydrodynamic equations are only the leading answer in the derivative expansion.  They receive an infinite series of corrections in higher derivatives, including nonlinear terms. Thermal and quantum fluctuations also modify these hydrodynamic equations. Both higher derivative and fluctuation corrections can be understood in the framework of an effective field theory (EFT) on the Schwinger--Keldysh (SK) contour~\cite{Grozdanov:2013dba, Harder:2015nxa, Crossley:2015evo, Haehl:2015uoc, Haehl:2018lcu, Jensen:2017kzi}; see~\cite{Liu:2018kfw} for a review. Modern numerical simulations are accurate enough to reveal these corrections~\cite{Michailidis:2023mkd} and experimentally it is possible to observe nonlinear response~\cite{Nicoletti_2016,chaudhuri2022}.

\begin{floatingfigure}[r]{0.4\textwidth}
\begin{center}%
   \includegraphics[width=0.2\textwidth]{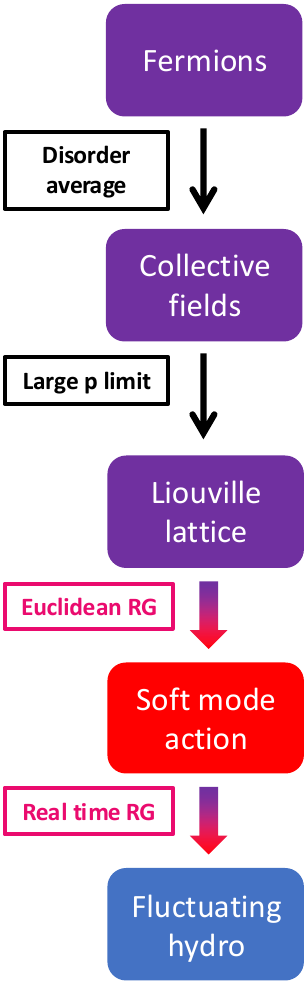}%
\caption{The logical flow of the derivation of the fluctuating hydrodynamics EFT from the large-$p$ SYK model.       \label{fig:logic}}%
  \end{center}%
\end{floatingfigure}

The SK EFT comes with undetermined transport coefficients at every order in the derivative expansion that need to be matched to the microscopic description. There are only a handful of models where higher derivative corrections can be computed analytically. One approach is based on starting with the Boltzmann equation and then applying gradient expansion to it~\cite{chapman1990mathematical,Denicol:2012cn}. While, conceptually, the transport coefficients can be determined in perturbative thermal quantum field theory, the computations are very challenging in practice~\cite{Jeon:1994if}. A class of examples where progress can be made, are large-$N$ gauge theories  with a holographic dual (e.g.~$\mathcal{N}=4$ super Yang--Mills theory). The fluid/gravity correspondence~\cite{Janik:2005zt,Bhattacharyya:2007vs,Bhattacharyya:2007vjd}  uncovers how the hydrodynamic degrees of freedom are encoded in the spacetime geometry and allows for the computation of transport coefficients to arbitrary order. In practice, however, these gravity computations become very complicated at subleading orders in derivatives~\cite{Baier:2007ix,Grozdanov:2015kqa}.\footnote{The fluid/gravity formalism is most developed at the level of the equation of motion. Capturing fluctuations requires complex geometries dual to the SK contour~\cite{Glorioso:2018mmw,Jana:2020vyx,Loganayagam:2022zmq}.}

\begin{figure}[h]
    \centering
    \includegraphics[scale=0.7]{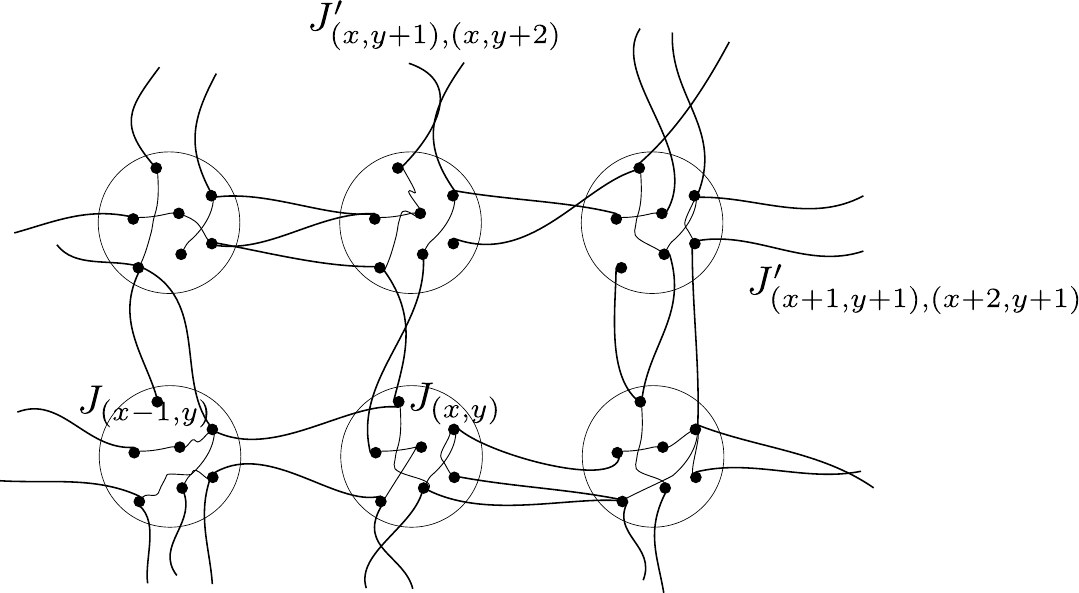}
    \caption{An illustration of the 2d SYK lattice. It is made from SYK dots, each containing $N$ Majorana fermions.  Independent on-site disorder coupling tensors $J_{(x,y)}$ and independent inter-site coupling tensors $J'_{(x,y),(x + 1, y )}$ and $J'_{(x,y),(x, y+1 )}$ are drawn from Gaussian distributions. }
    \label{fig:syk_lattice}
\end{figure}

The goal of this paper is to present a model where the SK EFT can be derived in a simple way starting from first principles. The model is the lattice generalisation~\cite{Gu:2016oyy} of the Sachdev-Ye-Kitaev (SYK) model~\cite{Sachdev_1993,KitaevTalks,Maldacena:2016hyu,Polchinski:2016xgd}. This model couples a collection of individual SYK dots, each containing $N$ Majorana fermions, into a lattice; see Figure~\ref{fig:syk_lattice}. We work in the strict large $N$ limit and focus on the low energy sector. We will primarily work with a one-dimensional SYK chain with nearest neighbour interactions for simplicity. However, our results are easily generalisable to higher-dimensional lattices or to examples with long-range hopping. Earlier works analysing energy transport in this model in the SK framework include~\cite{Zanoci:2021xyq,Zanoci:2022psr,Abbasi:2021fcz}: these works mainly focused on the linearised regime, whereas our results are nonlinear.

Besides obtaining the SK EFT with all transport coefficients (including nonlinear ones) determined for the SYK lattice, our derivation uncovers how hydrodynamic variables are embedded into the microscopic degrees of freedom. The most complete picture of this is achieved in the large-$p$ SYK lattice. Starting from the disordered fermion description, disorder averaging leads to a large-$N$ collective field description, which in the large-$p$ limit becomes a lattice of Liouville theories. Following earlier work~\cite{KitaevTalks,Maldacena:2016hyu,Kitaev:2017awl}, it was shown through a Euclidean RG computation in~\cite{Bucca:2024zrx,Berkooz:2024ifu} that a non-local soft mode action dominates the physics at low temperatures. In this paper, we show that for long Lorentzian times and for long wavelengths this non-local action can be turned into the local action of the SK EFT. See Figure~\ref{fig:logic} for the graphical presentation of the logical flow.

Briefly, the starting point of our derivation is eq.~\eqref{eq:final_action},
which is the effective low-energy action for the SYK chain for weak inter-site coupling $g$. This action has appeared in the literature before~\cite{Maldacena:2018lmt,Altland:2019lne,Almheiri:2019jqq,Bucca:2024zrx}; we provide novel arguments for it and discuss its regime of validity in detail. This action generalises the Schwarzian action for a single SYK dot. 
Akin to the single dot case, the lattice develops an approximate reparametrisation symmetry which governs the physics at low energies. However, in our case the reparametrisation $f_x(t)$ depends on both time $t$ and space $x$. The action has a local\footnote{This is because each site is a single undeformed SYK model. More complicated coupled SYK models are sometimes governed by a non-local action instead of the Schwarzian~\cite{Maldacena:2016upp,Milekhin:2021cou}.} contribution in terms of the familiar Schwarzian, plus a non-local in time term which comes from the inter-site coupling. 

Starting from this effective action, we implement a hydrodynamic expansion, which is a valid description when both time and space variations are small. Similarly to the single SYK case, the reparametrisations govern the physics when the time scale is bigger than the inverse coupling $1/J$, which is parametrically smaller than the local thermalisation time of order inverse temperature $\beta$ (in the regime $J \beta \gg 1$). It is for this reason that we can obtain an all-order expansion in time derivatives.
We recover diffusive transport of energy and compute subleading corrections, including non-Gaussianities in the fluctuations.
We also discuss the various symmetries of this model: space and time translation symmetries, conformal $\SL(2,\mathbb{R})$ reparametrisation symmetry, and the \emph{dynamical} Kubo-Martin-Schwinger (KMS) symmetry, which is the non-equilibrium incarnation of the microscopic time reversal symmetry. Note that the naive time reversal symmetry is broken in the SK EFT due to dissipation. The KMS symmetry allows us to extract the entropy current for our model~\cite{Glorioso:2016gsa}. 
We also extract higher derivative corrections to the energy density and flux. We will point out that the energy flux, despite being a local operator both in space and time, becomes a \textit{non-local} operator in the SK EFT in the sense that it receives contributions from both forward and backward folds of the Lorentzian closed-time contour.

The dynamical KMS symmetry is an important feature of SK EFTs. It is a non-local discrete symmetry that couples the degrees of freedom on separate folds of the closed-time contour. Thus far, bottom-up symmetry-based constructions of SK EFTs are only able to implement the KMS symmetry in full generality in the classical regime, i.e.~when the characteristic frequency $\omega$ of fluctuations satisfies the hierarchy $\omega \ll 1/t_{\text{micro}} \ll 1/(\hbar\beta)$; see~\cite{Blake:2017ris, Jain:2026obh} for some recent progress on this front. Here $t_\text{micro}$ denotes some characteristic microscopic timescale that controls the hydrodynamic expansion, while $\hbar\beta$ denotes the quantum timescale.\footnote{In large-$N$ systems $1/N$ controls fluctuations, and $1/N$ hence plays a role similar to $\hbar$. Here we distinguish the two. We call $\hbar \beta$ the quantum time scale, but it is not the scale at which  $1/N$ suppressed fluctuations around the saddle point become important. We discuss them separately in the main text.} By contrast, the SK EFT for energy diffusion derived in this work respects the full quantum version of the KMS symmetry and thus has a larger regime of applicability, $\omega \ll 1/t_{\text{micro}}\sim1/(\hbar\beta)$.

Furthermore, fluctuations in our theory scale as $(k \beta J/N)^{1/2}$, where $k$ is the wave vector of fluctuations. As mentioned above, the SK EFT derived in this work is valid for $\beta J\gg 1$, $N\gg 1$, and $k\ll1$. Due to the competition between these scales, it follows that our theory is also valid for large $O(1)$ fluctuations, provided that $\beta J$ is large enough to compensate the suppression from large $N$ and small $k$. In this regime, our theory reliably describes large non-equilibrium deviations away from the thermal state.

The paper is organised as follows: In Section~\ref{sec:micro} we discuss the microscopics of our model and give our derivation for the full effective action~\eqref{eq:final_action}.
Subsequently, in Section~\ref{sec:hydro} we obtain its hydrodynamic expansion. In Section~\ref{sec:energy} we derive the expressions for the energy density and flux, which contain both local and non-local pieces in terms of the dynamical variables of the model. In Section~\ref{sec:Noether} we discuss the continuous symmetries of the model and of the SK EFT, and derive the associated Noether currents.
In Section~\ref{sec:dynKMS} we discuss the (discrete) dynamical KMS symmetry of the action and its consequences: the fluctuation-dissipation theorem and the existence of an entropy current that satisfies the local second law of thermodynamics. Finally, in Section~\ref{sec:bottomup} we map our variables to the ones conventionally used in the SK EFT of energy diffusion and find a perfect match between the structures of the two theories, with all transport coefficients and higher derivative corrections fixed by our procedure.
In Section~\ref{sec:outlook} we provide an outlook by discussing numerous open questions.
Various technical details are delegated to the appendices and to a supplementary Mathematica notebook available with this \texttt{arXiv} submission and in the GitHub repository~\cite{github_syk_hydro_2026}.

\section{The nonlinear action of pseudo-Goldstone bosons}
\label{sec:micro}

Instead of the full SYK lattice, we start from the simpler case of the one-dimensional chain. In Appendix~\ref{app:kernel_shift} we discuss the four-point function and show that the action only depends on the underlying lattice Laplacian, allowing us to generalise our chain results to higher dimensions and to possible non-local spatial connectivity. 

The SYK chain consists of a chain of coupled SYK sites, with nearest neighbour interactions~\cite{Gu:2016oyy}.
Schematically, the Hamiltonian has the form:
\beq
H = \sum_{x \in \text{sites}}
\bigg( H^{\text{SYK-}p}_{x} + H^{\rm disordered \
 int}_{x, x+1} \bigg)\,. 
\eeq
Importantly, we take all the disorder tensors to be independent and consider the case of marginal inter-site coupling, this means that $H^{\rm disordered\ int}_{x,x+1}$ couples $p \times h$ fermions from the site $x$ and $p \times (1-h)$ from the site $x+1$.\footnote{We require $p \times h$ to be an integer.}  Explicitly:
\es{eqn:SYK chain hamiltonian}{
    H
    = \sum_{x=-M}^{M-1}
    \biggl(&\sum_{1\leq i_1< i_2< ... < i_p \leq N}J_{x|i_1i_2...i_p}\Psi_{i_1,x}...\Psi_{i_p,x} \\
    &\hspace{-3em}
    + \sum_{\substack{1\leq i_1< i_2< ... < i_{ph} \leq N\\ 1\leq j_1< j_2< ... < j_{p(1-h)} \leq N}}J'_{x,x+1|i_1...i_{hp}j_1...j_{p(1-h)}}\Psi_{i_1,x}...\Psi_{i_{hp},x}\Psi_{j_1,x+1}...\Psi_{j_{(1-h)p},x+1}\biggr)\,,
}
where we identify the lattice site at $x=\pm M$.
The coupling constants $J_x$ and $J'_{x,x+1}$ are drawn from Gaussian distributions with zero mean and variances given by
\es{}{
    \left\langle J^2_{x|i_1,..i_p}\right\rangle 
    &= \frac{(p-1)!}{N^{p-1}}J_0^2\,,\\
    \left\langle J'^2_{x,x+1|i_1...i_{hp}j_1...j_{(1-h)p}}\right\rangle 
    &= \frac{(ph)!(p(1-h))!}{pN^{p-1}}J_1^2\,,
}
for constants $J_0$ and $J_1$. It is convenient to introduce
\es{eqn:constants for the chain}{
\mathcal{J}^2_{0,1} &\equiv\frac{p}{2^{p-1}}J^2_{0,1}\,, \qquad
\mathcal{J}^2\equiv\mathcal{J}_0^2+\mathcal{J}_1^2\,, \\
\coup &\equiv \frac{4J_1^2\, b^p}{8p}\,, \hspace{1.3cm} b^p = \frac{1}{\pi (J_0^2 + J_1^2)} \l( \frac{1}{2} - \frac{1}{p} \r) \tan \l( \frac{\pi}{p} \r)\,,
}
with $g$ the effective inter-site coupling that we will often use throughout the paper. With this, the final form of the low-energy effective action in Euclidean time is~\cite{Maldacena:2018lmt,Altland:2019lne,Almheiri:2019jqq,Bucca:2024zrx}:\footnote{Note that~\eqref{eq:final_action} corrects a factor of $1/4$ mistake in~\cite{Bucca:2024zrx}.} 
\es{eq:final_action}{
    I_E[f]=
    -N \sum_{x=-M}^{M-1}
    &\Bigg[ \frac{\alpha_S}{\Jc} \int_0^{\beta} d\tau\, \Sch\!\big[\tan(f_x/2),\tau\big] \\
    &\hspace{-2em}
    + \frac{\coup}{4}\int_0^\beta d\tau_1 d\tau_2
    \l(  \frac{ f_x'\left(\tau _1\right) f_x'\left(\tau _2\right)}{\sin ^2\left(\frac{ f_x\left(\tau _1\right)-f_x\left(\tau
   _2\right)}{2}\right) } \r)^{h}  \l( \frac{ f_{x+1}'\left(\tau _1\right)  f_{x+1}'\left(\tau _2\right) }{ \sin ^2\left(\frac{ f_{x+1}\left(\tau _1\right)-f_{x+1}\left(\tau
   _2\right)}{2}\right) } \r)^{1-h} 
   \Bigg]\,,
}
where $\alpha_S$ is a dimensionless constant\footnote{$\alpha_S=1/(4p^2)$ at large $p$, and its value can only be computed numerically at general $p$~\cite{Maldacena:2016hyu}.}  and primes denote derivatives with respect to the time argument.  Furthermore, $\Sch$ denotes the Schwarzian derivative
\begin{align}
    \Sch[h(t),t] 
    = \p_t\left(\frac{\p_t^2h}{\p_th}\right)
    - \frac12 \left(\frac{\p_t^2h}{\p_th}\right)^2\,.
\end{align}
The derivation assumes $g\lesssim \alpha_S/(\beta\Jc)$.

\pagebreak
Beyond the large-$p$ SYK chain in the stated parameter range, the action is valid for any  $p \ge 4$ in two completely different regimes:
\begin{itemize}
    \item For small $J_1/J$, but to arbitrary order in $x$-derivatives~\cite{Bucca:2024zrx}.
    \item For arbitrary $J_1/J$, but to quadratic order in the  $x$-derivative expansion.
\end{itemize}
The latter follows from the fact that for an $x$-independent $f_x(\tau)$ this term becomes a total derivative, hence $f_x(\tau)$ slowly varying in space will give a small action irrespective of $J_1/J$.
We corroborate both of these statements by showing that the linearisation of this action correctly reproduces the four-point function in Appendix~\ref{app:kernel_shift}. There  we also demonstrate that the action only depends on the lattice Laplacian.

\section{Long wavelength expansion}
\label{sec:hydro}

In this section, we extend the non-local Euclidean effective action~\eqref{eq:final_action} of the SYK chain onto the Schwinger--Keldysh closed-time contour and perform a long  wavelength expansion to arrive at the corresponding local SK EFT. The classical equation of motion arising from the SK EFT  recovers energy diffusion complete with higher derivative corrections.

\subsection{Local action from the non-local action}
\label{sec:loc_from_nonloc}

\begin{figure}[h]
     \centering
\includegraphics[scale=0.7]{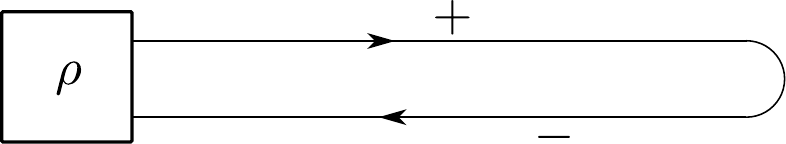}
     \caption{The SK contour and the initial density matrix $\rho$.}
     \label{fig:SK}
 \end{figure}

 We analyse the theory on the Schwinger--Keldysh (SK) contour with long Lorentzian folds. Under Heisenberg evolution, an operator $\Oc$ evolves as $ e^{+i H t} \Oc e^{-i H t}$. The two evolution operators on the left and right, when converted into a path-integral, correspond to forward ``$+$'' evolution and ``$-$'' backward evolution; see Figure~\ref{fig:SK}. To make the evolution operators well-behaved, one adds $\pm i \epsilon$ to the different sides of the contour. 

The soft mode action is non-local in time. On the Lorentzian folds of the SK contour, the non-locality is only on the time scale $\beta$.  Hence we expect that a local action emerges at scales longer than $\beta$: this is the hydrodynamic regime.
To see this, let us consider the saddle point
\es{fSaddle}{
f_x^*(\tau)={2\pi i\ov \beta}\, t\,,
}
where $t = -i\tau$ is the Lorentzian time. By plugging the saddle into~\eqref{eq:final_action} with $h=1/2$, we encounter the term 
\es{sinh}{
\frac{ \sqrt{f_x'\left(\tau _1\right)} \sqrt{f_x'\left(\tau _2\right)}}{\sqrt{\sin ^2\left(\frac{ f_x\left(\tau _1\right)-f_x\left(\tau
   _2\right)}{2}\right)}}=-{2\pi i\ov \beta}\, 
   {1\over \sinh\le({\pi \abs{t_{1}-t_2}/ \beta}\ri)}\,,
}
which is indeed exponentially localised for time separations of $|t_{1}-t_2|\sim \beta$. 

Armed with this understanding, we can set up the long wavelength expansion in the following manner:
\begin{itemize}
\item We introduce $\lam$ as a formal small expansion parameter, and we scale
\es{scaling}{
t\sim\lam^{-2}\,,\qquad x\sim \lam^{-1}\,.
}
This scaling is inspired by the expected diffusive scaling. (To apply the final action to describe energy transport, one can set $\lam=1$.)
    
\item We go to the average-difference basis in time, and choose the average time to be large, while keep the difference time $O(\lam^0)$: 
\es{CM}{
t_\pm={s\ov \lam^2}\pm{u\ov 2}\,,
}
where $t_\pm$ denote the time coordinates on the forward and backward parts of the contour.

\item We go to an average-difference basis in the space of fields as well. On-site we introduce the usual classical and noise (or quantum) fields that we denote by $F$ and $Q$. Furthermore, to facilitate going to the continuum description in space, we also introduce $G$ and $R$ for the discretised gradients of $F$ and $Q$ respectively. We use {\textcolor{red}{red $(\pm)$}} superscripts to denote contour indices and write:
\es{FieldCM}{
f_{x\pm \frac12}^{{\color{red}(\pm)}}(it) 
&= {2\pi i \ov \lam^2} F_{x\pm \frac12}^{{\color{red}(\pm)}}(\lam^2 t) \\
&= 2\pi i\left[{F_x(\lam^2 t)\ov \lam^2}
\pm {G_x(\lam^2 t)\ov 2 \lam}
{\color{red}\,\pm\,} 
\left( {Q_x(\lam^2 t)\ov 2}
 \pm{\lam R_x(\lam^2 t)\ov 4 } \right)\right]\,,
}
where the $i$ prefactor is to implement the Wick rotation and $x$ now stands for the midpoint between lattice points. The $\lam^2 t$ argument of functions implements the gradient expansion via $\p_t^n F=\lam^{2n} F^{(n)}=O(\lam^{2n})$. 

The grading of $F$ with $\lam^{-2}$ follows from the form of the saddle point~\eqref{fSaddle}. The grading of $G$ is down by $\lam$ compared to $F$ because of the suppression of spatial gradients. The same applies to the grading of $R$ relative to $Q$. Hence the only unfixed grading is that of $Q$ itself. From~\eqref{Ipm1} below, it is clear that $Q_x$ has to be smaller than or equal to $\lambda^0$. It is the dynamical KMS symmetry that fixes the  grading of $Q_x$ to be  $\lambda^0$, as we discuss later on in the paper. The $\lam$ scaling prescribed here will then lead to an action that is $O(\lam^0)$.

\item We plug~\eqref{FieldCM} and~\eqref{CM} into~\eqref{eq:final_action} and expand in $\lam$. Crucially, the $u$ and $s$ integrals factorise at every order in $\lam$, and the $u$ integrals can be performed, giving rise to a local action in the average time $s$.

\item Finally we go to the spatial continuum limit, where we introduce two continuum fields. The discrete fields at lattice site $x=y/\lam$ are related to the continuum fields through
\es{SpatialCont}{
F_{ y/\lam}(s)&={F(s,y+\lam/2)+F(s,y-\lam/2)\ov 2}\,, \qquad
G_{ y/\lam}(s)={F(s,y+\lam/2)-F(s,y-\lam/2)\ov \lam}\,,\\
Q_{ y/\lam}(s)&={Q(s,y+\lam/2)+Q(s,y-\lam/2)\ov 2}\,, \qquad
R_{ y/\lam}(s)={Q(s,y+\lam/2)-Q(s,y-\lam/2)\ov \lam}\,,
}
where $G$ and $R$ are manifestly discretised derivatives.\footnote{A slightly simpler choice would be to take $F_{ y/\lam}(s)=F(s,y)$ which differs from the choice we made at $O(\lam^2)$ by $\p_{yy} F$. This latter choice would slightly simplify a few subsequent formulas, but would make coupling to a background geometry, that we use to extract the energy current, more subtle.}

\end{itemize}

We illustrate this procedure in a low order computation; the computation goes the same way at every order, and can be automated straightforwardly in Mathematica; our code is available both as attached to the \texttt{arXiv} submission and in the GitHub repository~\cite{github_syk_hydro_2026}. 
In most of the paper we focus on the $h=1/2$ case. Extending to $h \neq 1/2$ is a simple exercise, included in Appendix~\ref{app:hdep}. We found that it results in parity-breaking fluctuation terms starting at $O(\lam^2 Q^2)$ in the action, but does not change the equation of motion up to $O(\lam^3)$. Similarly, one can analyse other lattices with possibly non-local connectivity.

\subsubsection*{Schwarzian action}

We focus on the long Lorentzian time sections of the SK contour, and correspondingly convert to the conventions for Lorentzian action: $I_E=-iI$. For the Schwarzian part of the action, Wick rotation yields
\es{eq:sch-lorentzian-action}{
    I_{\Sch}^{\pm} 
    &= \frac{N \alpha_S}{\Jc} \sum_{x=-M}^{M-1} 
    \int d t\, \Sch\!\big[\tan(f^{\pm}_x(it)/2),it\big] \\
    &= -\frac{N \alpha_S}{\Jc} \sum_{x=-M}^{M-1} 
    \int d t\, \Sch\!\big[\tanh(\pi F^{\pm}_x\l(\lambda^2t\r)/\lambda^2),t\big]\,,
}
with one copy residing on each fold of the Lorentzian contour. 
Combining the action on the two Lorentzian folds, we get the Schwarzian contribution to the SK effective action
\es{eq:sch-lorentzian-twofolds}{
    I_{\SK,\Sch}
    &= I_{\Sch}^+ - I_{\Sch}^- =
    \frac{4 \pi^2 N\alpha_S}{\Jc} 
    \sum_{x=-M+1/2}^{M-1/2}
    \int d s\,F'_{x}(s) Q'_{x}(s) +O(\lam^2)\,,
}
where we changed to the rescaled time variable $s$.
Note that since time runs backwards on the backward fold of the contour, the respective contribution to the SK action appears with a minus sign.
In going from (\ref{eq:sch-lorentzian-action}) to (\ref{eq:sch-lorentzian-twofolds}), we have implemented a variable change from $t_\pm \to s/\lambda^2$ independently for both the Lorentzian folds and we have dropped a total derivative term. 
Passing onto the continuum limit, we find:\footnote{Because we defined the continuum fields to be linear combinations of fields on neighbouring lattice sites in~\eqref{SpatialCont}, at higher order in $\lam$ the Lagrangian ${\cal L}_{\SK,\Sch}$ will contain space derivatives. However, these are always total derivatives and can be freely dropped.}
\es{eq:Isch}{
    I_{\SK,\Sch}
    = \frac{4 \pi^2 N\alpha_S}{\Jc\lambda}
    \int d s dy\,\dow_s F(s,y) \dow_s Q(s,y) +O(\lam^2)\,.
}

\subsubsection*{Inter-site term}

For the non-local inter-site interaction term, Wick rotation yields an action on each fold of the Lorentzian contour.
Since both time integrals can run over both Lorentzian folds, the resulting SK action is a sum of four terms
\es{Itot}{
I_{\SK,\text{non-loc}}
= I_{++} - I_{+-} - I_{-+} + I_{--}\,.
}
We pick the off-diagonal term $I_{+-}$ for illustration. After passing onto $s$ and $u$ variables, the leading contribution to $I_{+-}$ at small $\lam$ is given as\footnote{The $i\ep$ shift was originally in the time argument of the $F^{-}$ functions. We series expand in $\ep$ to get an $iF'(s) \ep$ shift, which we can replace by $i\ep$, since the $F$ functions are required to be monotonic in time. }
\es{Ipm1}{
-I_{+-}=i\coup N \sum_{x=-M+1/2}^{M-1/2}
\int ds du \, {\pi^2 F_x'(s)^2\ov \lam^2\, \sinh^2\le(\pi\le(F_x'(s)u+Q_x(s)\ri)+i\ep\ri)} +O(\lam^0)\,.
}
By the shift and rescaling $u=(\tilde u -\pi Q_x(s))/(\pi F_x'(s))$, we can evaluate the $u$ integral:
\es{Ipm2}{
I_{+-}&=i\coup N
\sum_x
\int ds \int d\tilde u \, {\pi F_x'(s)\ov \lam^2\, \sinh^2\le(\tilde u+i\ep\ri)} +O(\lam^0)\\
&=-i\coup N\sum_x\int ds \, {2\pi F_x'(s)\ov \lam^2} +O(\lam^0)\,.
}
This term is not interesting: it is a total derivative term $F'(s)$. Furthermore, it turns out to cancel with contributions from the other fold choices in~\eqref{Itot}. However, the same change of variable to $\tilde u$ also simplifies the interesting higher order terms. 

Relegating details to the supplementary Mathematica notebook, to leading order in $\lam$ we get the following effective action
\es{Lint}{
I_{\SK,\text{non-loc}}
= {2\pi^3 \coup N \ov 3}\sum_x\int ds
\le[\le({Q_xG_x'^2\ov F_x'}-R_xG_x'\ri)+i{\le(R_x F_x'-Q_x G_x'\ri)^2\ov F_x'}\ri]+O(\lam^2)\,.
}
We can go to the spatial continuum limit, to obtain:
\es{Lint2}{
I_{\SK,\text{non-loc}}
={2\pi^3 \coup N\ov 3\lambda}
\int ds dy
\le[\le(
{Q (\p_{sy} F)^2\ov \p_s F}-\p_yQ \p_{sy}F\ri)+i{\le(\p_y Q \p_s F-Q \p_{sy}F\ri)^2\ov \p_s F}\ri]+O(\lam^2)\,.
}

We can combine the contribution~\eqref{eq:Isch} from the Schwarzian and interactions above to arrive at the total SK effective action
\es{Leff2}{
{I}_\text{SK}
&= \coup N{2\pi^3\ov 3\lambda}\int dt dx
\Bigg[
{1\ov D}  \p_t F \p_tQ +{Q (\p_{tx} F)^2\ov \p_t F}-\p_xQ \p_{tx}F
+ i{\le(\p_x Q \p_t F-Q \p_{tx}F\ri)^2\ov \p_t F}+O(\lam^2)\Bigg]\,,
}
where we have denoted the rescaled time and space variables with $(t,x)$ for ease of reading and introduced the diffusion constant 
\beq
\label{eq:diffusion_c}
D= {2\pi^3 \coup\ov 3}\Big/\le(4 \pi^2 \alpha_S\ov\Jc\ri)\,.
\eeq
This diffusion constant exactly reproduces the one obtained in~\cite{Gu:2016oyy,Choi:2020tdj} by linearising around the thermal state. 
The effective Lagrangian is given to $O(\lam^2)$ in Appendix~\ref{app:highorder} and to $O(\lam^4)$ in the Mathematica file.


\subsection{Energy diffusion}
\label{sec:energy-diff}

Every term in the effective action~\eqref{Leff2} is at least linear in the noise field $Q$, even at higher orders in $\lambda$. This is simply because the SK effective action is defined via contributions from the forward Lorentzian fold minus those from the backward fold, hence it must vanish when the noise field $Q$ is turned off. As a consequence, extremising the effective action with respect to $F$ leads to a classical equation of motion for $Q$ that is identically satisfied by the noiseless configuration $Q = 0$. By contrast, extremising the effective action with respect to $Q$ leads to an equation of motion for $F$, which upon setting $Q=0$ takes the form:
\begin{align}
    \p_{t}^2 F 
    = D\p_{xxt}F
    + D\frac{(\p_{xt}F)^2}{\p_t F}
    + O(\lambda^2)\,.
\end{align}
To understand the resulting hydrodynamic theory, we identify the temperature with $T=\p_t F$, which leads to the heat equation
\es{DiffEq}{
\p_t T= D\le(\p_{x}^2 T+{(\p_x T)^2\ov T}\ri) 
+ O(\lambda^2)
\,. 
}
From the thermodynamics of the Schwarzian, we know that the energy density ${\cal E}^t$ is proportional to $T^2$ at leading order in $\lambda$. In terms of this, we recover the conventional form of the energy diffusion equation:
\es{DiffEqEn2}{
\p_t \,{\cal E}^t= D\,\p_{x}^2\, {\cal E}^t 
+ O(\lambda^2)
\,. 
}
We will discuss this is detail in Section~\ref{sec:energy}.

Fluctuations around the classical hydrodynamic regime are small in our theory, as they are suppressed by $1/N$ that acts as an effective $\hbar$. This is most easily seen by rescaling the noise variable $Q=\tilde Q/\sqrt{N}$, in which case the Lagrangian takes the form
\es{Leff4}{
{I}_\text{SK}
&= {2\pi^3\coup \ov 3\lambda}\int dt dx
\left[
{-}\sqrt{N} \tilde Q \le( \frac{\p_{t}T}{D}
    - \p_{x}^2T
    - \frac{(\p_{x}T)^2}{T}
\ri)
+ i{\le(T\p_x \tilde Q -\tilde Q \p_{x}T\ri)^2\ov T}
\right]+O(\lam^2)\,,
}
up to a total derivative term.
The fluctuations of $\tilde Q$ are   $\langle\tilde Q^2\rangle=O(N^0)$ in the path integral. Therefore at leading order in $N$, the variable $\tilde Q$ merely acts as a Lagrange multiplier to impose the equation of motion of $F$. At subleading order in $1/N$, the equation of motion acquire noise contributions. We will return to these in Section~\ref{sec:stochastic} after having discussed the energy current.

\subsection{Scaling of stochastic and quantum fluctuations}

To better understand the role of fluctuations in our EFT, let us expand the effective action~\eqref{Leff2} around the classical equilibrium solution: $F=t/\beta$ and $Q=0$, i.e.~we take
\es{SaddleLin}{
F(t,x)={t\ov \beta}+\epsilon \,f(t,x)\,,\qquad Q(t,x)=\epsilon \,q(t,x)\,,
}
where $\epsilon$ is a symbolic expansion parameter that we set to one below. At linear order in fluctuations, the effective action is merely a total derivative. At quadratic order, we find the free Gaussian action
\es{}{
{I}_\text{SK,2}
&= \coup N{2\pi^3\ov 3\beta\lambda}\int dt dx
\Bigg[
{1\ov D} \p_t q\,(\beta\p_{t} f)  - \p_x q\,\p_{x}(\beta\p_t f)
+ i(\p_x q)^2 + O(\lam^2)
\Bigg]\,,
}
The action is dimensionless (in units where $\hbar=1$). This is a Gaussian action, so we can immediately read off the size of the $q$ and $f$ fluctuations:
\begin{align}
    q(t,x) \sim \beta \dow_t f(t,x)
    \sim \sqrt{\lambda k\frac{\beta \Jc}{\alpha_S N}}
    \equiv Z,
\end{align}
where $k$ is the wave vector of fluctuations and we have used the definition of $D$ from~\eqref{eq:diffusion_c}.\footnote{In this analysis we have assumed diffusive scaling $\om\sim D k^2$ for the fluctuations, which may not be the only regime of interest for a particular observable. We have also assumed that fluctuations are Gaussian; this does not apply for $g=0$, where we get a non-dissipative theory of decoupled SYK dots. A toy version of the resulting integral is $\int dq df\, e^{i N q f}$, for which it is quite subtle how to define the size of fluctuations: see page 7 of~\cite{Stanford:2021bhl} for an insightful analysis of this elementary integral. }

We can make some important inferences from here. First, as in the case of a single SYK dot, all fluctuations around the thermal state are suppressed at large $N$ as long as $\beta$ is not too large.\footnote{This is strictly true for the partition function. Studying observables is more subtle, because they introduce additional scales. For example, real-time two-point functions in this theory behave as $G \sim 1/\sinh^{2\Delta}(\pi t/\beta)$. Fluctuations of the inverse temperature $\beta= 1/\pr_t F$ lead to the fluctuations in $G$ which scale as $\delta G/G \sim t^2 \bra  (\pr_t f)^2\ket $ and become large at late times. The same mechanism is operational in the single SYK dot.} Second, fluctuations are also suppressed in the hydrodynamic regime $k\to 0$, irrespective of the magnitude of $N$. This has a simple physical explanation: a sector with non-zero $k$ roughly involves $1/k$ individual SYK dots leading to an effective $N_\text{eff} \sim N/k$. In higher spatial dimensions, the fluctuations analogously scale as $\sim k^{d/2}$. Furthermore, the fluctuations of the noise field $Q$ are suppressed by a $\beta\omega$ factor relative to $F$, meaning that the noise becomes less important at large temperatures or small frequencies, which is also a standard hydrodynamic feature.

The SK EFT for the SYK chain that we have derived also assumes $\beta\Jc\gg 1$. This results in a competition of scales in the fluctuations. In particular, fluctuations are only suppressed if the ratio $Z^2 = \lambda k\,\beta\Jc/N$ is sufficiently small for a given wave vector. Interestingly, our theory is also valid for large fluctuations with $Z^2\sim O(1)$, provided that $\beta\Jc$ is large enough to overcome the suppression from $1/N$ and $k$. In this regime, our theory will reliably describe large non-equilibrium deviations away from the thermal state.

In the regime where $Z\ll 1$, we can use our SK EFT to set up a systematic perturbative expansion around the thermal state. For example, at cubic order in fluctuations we find 3-point interactions
\es{}{
{I}_\text{SK,3}
&= \coup N{2\pi^3\ov 3\beta\lambda}\int dt dx
\Bigg[
(\beta\p_{tx} f)^2 q
+ i(\p_x q)^2 (\beta\p_t f)
+ iq^2 (\beta\p_{txx}f)
+ O(\lam^2)
\Bigg],
}
each of which scale as $Z$. In particular, we find that both the dissipative $qf^2$ as well the noise $q^2f$ interactions scale in the same manner. This is a manifestation of the fluctuation-dissipation theorem, which we will return to in Section~\ref{sec:fluctdiss}. Similarly at the level of quartic fluctuations, we find 4-point interactions
\es{}{
{I}_\text{SK,4}
&= \coup N{2\pi^3\ov 3\beta\lambda}\int dt dx
\Bigg[
{-} (\beta\p_t f)(\beta\p_{tx} f)^2 q
+ iq^2 (\beta\p_{tx}f)^2
+ O(\lam^2)
\Bigg],
}
which scale as $Z^2$. This construction parallels the recent studies of long-time tails in diffusive systems using SK EFTs in~\cite{Chen-Lin:2018kfl, Jain:2020zhu, Jain:2020hcu, Jain:2026obh}.

The above discussion can similarly be extended to the SK EFT at higher orders in $\lambda$. At $O(\lambda^2)$, we find that all $n$-point interactions scale as $\lambda^2 k^2 Z^{n-2}$. The explicit effective action at $O(\lam^2)$ is presented in Appendix~\ref{app:highorder}. In particular, the theory features non-Gaussian noise interactions at this order of the kind $q^3$, $q^3f$, $q^4$, etc., which scale in the same manner as their Gaussian noise $q^2f$, $q^2f^2$ or dissipative $qf^2$, $qf^3$ counterparts (at the same $\lam$ order). 

Starting at $O(\lam^4)$, we begin to see the effects of quantum fluctuations. To wit, we find two classes of $n$-point interactions at $O(\lam^4)$, schematically given as 
\es{sto_vs_quant}{
    \text{Stochastic:}&\quad  
    \coup N{2\pi^3\ov 3\beta\lambda}\int dt dx\,\lambda^4\p_x^6
    \bigg[ q^m (\beta\dow_tf)^{n-m} \bigg]
    \sim (\lam k)^4 Z^{n-2}, \\
    \text{Quantum:}&\quad  
    \left\{
    \begin{array}{l}
    \displaystyle
    \coup N{2\pi^3\ov 3\beta\lambda}\int dt dx\,\lambda^4 \beta^2 \dow_t^2\p_x^2
    \bigg[ q^m (\beta\dow_tf)^{n-m} \bigg] \\
    \displaystyle
    \coup N{2\pi^3\ov 3\beta\lambda}\int dt dx\,
    \frac{\lambda^4 \beta^2}{D} \dow_t^3
    \bigg[ q^m (\beta\dow_tf)^{n-m} \bigg] \\
    \end{array}
    \right\}
    \sim (\lam k)^4 Z^{n-2} (\beta D)^2,
}
for some $n\geq m\geq 1$. The notation above means that the derivatives are distributed in some way on the fields in the parenthesis. The dichotomy between stochastic and quantum terms comes from the observation that stochastic terms have extra factors of the temperature. The explicit effective action at $O(\lam^4)$ is given in the supplementary Mathematica notebook. 

Restoring $\hbar$ and the lattice spacing $a$, quantum interactions are suppressed with respect to the stochastic interactions by a factor of $(\hbar\beta/t_\text{micro})^2$, where 
\es{tmicro_def}{
t_\text{micro}={a^2\ov D} \propto  \frac{\alpha_S}{g\Jc}\,.
}
As discussed in the introduction, quantum fluctuations are suppressed in the classical regime $\omega\ll 1/t_\text{micro} \ll 1/(\hbar\beta)$, where classical SK EFTs of hydrodynamics are valid.
We emphasise that  quantum here only means the terms~\eqref{sto_vs_quant} and not $1/N$ loop  corrections.
By contrast, our theory is also valid in the regime where quantum and stochastic fluctuations are comparable, i.e.~$\omega\ll 1/t_\text{micro} \sim 1/(\hbar\beta)$. The classical domain hence corresponds to $g\ll\alpha_S/(\beta\Jc)$, while in the regime $g\sim\alpha_S/(\beta\Jc)$, which is the boundary of applicability of~\eqref{eq:final_action}, the quantum and stochastic fluctuations are equally important. See~\cite{Jain:2026obh} for related discussion in the SK EFT of charge diffusion.

\section{Conserved energy currents}
\label{sec:energy}

In this section we discuss energy conservation in our theory in detail. There are two variants of energy current in our theory: one living on each Lorentzian fold of the closed-time contour. In the classical limit, i.e.~in absence of the noise field $Q$, these energy currents coincide. However, more generally, they differ from each other and contain terms that couple the two folds together owning to the non-local interaction term in the soft mode action~\eqref{eq:final_action}. We will use the standard prescription of coupling to background fields to derive the energy currents. Later in Section~\ref{sec:Noether} we will revisit these currents from the perspective of time translation symmetries.

\subsection{Coupling to background fields}

In Lorentz-invariant theories, one can derive the energy-momentum tensor by coupling the theory to a background spacetime metric and varying the action with respect to the metric. The idea~\cite{landau1975classical} behind the derivation is that the coupled theory is invariant under spacetime diffeomorphisms. An infinitesimal diffeomorphism of the action consists of the variation over the dynamical fields (which vanishes on equations of motion) plus the variation over the metric. Imposing that this latter part is zero yields the conservation equation for the energy-momentum tensor. 

In our theory we can perform a similar exercise; see e.g.~\cite{Blake:2017ris, deBoer:2017ing, Armas:2020mpr}. However, instead of a metric, we couple the theory to the background einbein $e^\pm_{t}$ and the background connection $e^\pm_x$ such that the theory is invariant under (infinitesimal) time diffeomorphisms on each Lorentzian branch of the contour:\footnote{Compared to the work of~\cite{Blake:2017ris}, we have $e_t = e$ and $e_x = - e w_x$. In works on boost agnostic hydrodynamics~\cite{deBoer:2017ing, Armas:2020mpr}, the background fields $e_\mu$ are denoted as $n_\mu$ or $\tau_\mu$. To avoid potential confusion, note that $e_\mu$ is \emph{not} the Zweibein.}
\beq
\label{eq:diff}
    t_\pm \ra t_\pm 
    - \xi_\pm\left(t_\pm,x\right)\,,
\eeq
where $t_\pm \to t_\pm/\lambda^2$ and $x\to x/\lambda$ are understood as rescaled coordinates.
Note that $F^\pm(t,x)$ are scalars under time diffeomorphisms, i.e.~$F^\pm(t,x)\to F^\pm(t+\xi_\pm(t,x),x)$
To promote the time translation symmetries of our model to time diffeomorphisms, we introduce the time and space covariant derivatives of $F^\pm$ as
\es{Covariantise}{
    \pr_t F^\pm 
    &\ra D_t F^\pm = \frac{1}{e^\pm_{t}} \pr_t F^\pm\,, \\
    \pr_x F^\pm 
    &\ra D_x F^\pm  
    = \pr_x F^\pm  - \frac{e^\pm_x}{e^\pm_t} \pr_t F^\pm.
}
Correspondingly, we require that $D_t F^\pm$ and $D_x F^\pm$ also transform as scalars, which requires the background fields to transform as
\es{eq:background-field-transformation}{
    e_\mu(t,x)
    &\to e_\mu(t+\xi(t,x),x) 
    + \partial_\mu\xi(t,x)\, e_t(t+\xi(t,x),x) \\
    &= \Big(e_\mu + e_t \pr_\mu \xi
    + \xi\partial_t e_\mu\Big) (t,x)
    + O(\xi^2)\,,
}
where the index $\mu$ runs over $t$ and $x$ and we have suppressed the $\pm$ labels for clarity.
Note that $D_t$ and $D_x$ derivatives do not commute. To turn off the background fields and return to the original theory, we can set $e^\pm_t = 1$ and $e_x^\pm = 0$.

Consider that we have obtained the action $I_\SK$ of our theory appropriately coupled to background fields. Having done that, we define the energy current ${\cal E}^\mu_\pm$ on the two folds of the contour by varying with respect to the associated background fields\footnote{We have defined the background fields and energy current in terms of the $\lambda$-rescaled coordinates. We could define the respective unscaled versions
\es{}{
\hat e_t(t,x) = e_t(\lambda^2 t,\lambda x)\,, \quad 
\hat e_x(t,x) = \frac{1}{\lambda}e_x(\lambda^2 t,\lambda x)\,, \quad 
\hat\cE^t(t,x) = \cE^t(\lambda^2 t,\lambda x)\,, \quad 
\hat\cE^x(t,x) = \lambda \,\cE^x(\lambda^2 t,\lambda x)\,.
}
In terms of unscaled coordinates, the $\lambda^3$ factor in~\eqref{eq:energy-current} drops out as it cancels with the one arising from the integration measure $dt\, dx$.}
\begin{align}\label{eq:energy-current}
    {\cal E}^\mu_\pm = \mp \lambda^3\, \frac{\delta I_\SK}{\delta e^\pm_\mu}\,.
\end{align}
The right-hand side above is evaluated at $e^\pm_t = 1$ and $e_x^\pm = 0$.
The opposite signs in~\eqref{eq:energy-current} compensate the opposite sign of the time integral in the forward and backward folds of the contour. An infinitesimal variation of the action with respect to $\xi_{\pm}$ must be zero, which yields the following condition when the background fields are switched off, i.e.
\begin{align}
    0 = \delta_{\xi}I_\SK
    = \int dt dx  
    \left(\underbrace{\frac{\delta I_\SK}{\delta F^+}}_{\rm EoM} 
    \xi_+ \partial_t F^+
    + \underbrace{\frac{\delta I_\SK}{\delta F^-}}_{\rm EoM} 
    \xi_- \partial_t F^-
    - \frac{1}{\lambda^3} {\cal E}^\mu_+ \pr_\mu \xi_+
    + \frac{1}{\lambda^3} {\cal E}^\mu_- \pr_\mu \xi_-
    \right)\,.
\end{align}
The resulting energy conservation equations on the two folds are given as\footnote{In the presence of background fields, the energy conservation equation gets modified by a power term
\begin{align}\label{eq:conservation-sourced}
    \dow_\mu\cE^\mu = -\cE^x E_x\,,
\end{align}
on each fold of the contour. Here $E_x = (\dow_x e_t - \dow_t e_x)/e_t$ is the analogue of the background electric field for $e_\mu$. In relativistic theories, $E_x = -\Gamma^x_{tt}/g_{tt}$ is a component of the Christoffel connection upon identifying the spacetime metric as $g_{\mu\nu} = -e_\mu e_\nu + \delta_{\mu x}\delta_{\nu x}$.
}
\begin{align}
    \partial_\mu {\cal E}^\mu_\pm
    = 0.
\end{align}

\subsection{Non-local action with background fields}
\label{sec:coupling-action-to-background}

Let us start by coupling the Schwarzian part of the action to background fields and rederive the standard fact that the energy of the Schwarzian theory is proportional to the Schwarzian itself. 
We write the Schwarzian action in the continuum limit in the presence of background fields by promoting~\eqref{eq:sch-lorentzian-action} as 
\es{eq:sch-lorentzian-action-e}{
    I_{\Sch} 
    &= -\frac{\lambda N \alpha_S}{\Jc} 
    \int d t dx\, e_t(t,x)\Sch\!\big[\tanh(\pi F\l(t,x\r)/\lambda^2),t\big]_D,
}
where subscript $D$ indicates that all derivatives are covariant. Explicitly:
\begin{align}
-e_t \Sch\le[\tanh(\pi F/\lambda^2),t\ri]_D 
&= -\frac{1}{e_t} \Sch\le[\tanh(\pi F/\lambda^2),t\ri]  
+ \frac{\p_{tt}e_t}{e_t^2} 
- \frac{3 (\p_t e_t)^2}{2e_t^3}\,,
\end{align}
where we have dropped the $\pm$ labels for clarity. The variation with respect to $-e_t$ at $e_t=1$ is exactly minus the Schwarzian, which is the energy density. To wit
\begin{align}
    {\cal E}^t = - \frac{\lambda^4 N \alpha_S}{\Jc}
    \Sch\le[\tanh(\pi F/\lambda^2),t\ri]\,,
\end{align}
on each fold of the contour. Note that the Schwarzian part of the action does not involve $e_x$, therefore the corresponding energy flux is exactly zero.

Introducing background fields to the non-local coupling term of the action is slightly more subtle. Firstly, let us write down the continuum version of the inter-site coupling term
\es{}{
    I_{\text{non-loc}} =
    -\frac{i\pi^2 N\coup}{\lambda^5} 
    &\int d t_1 d t_2 dx
    \l(  \frac{ \dow_{t_1}F\left(t_1,x-\lambda/2\right) \,
    \dow_{t_2}F\left(t_2,x-\lambda/2\right)}
    {\sinh ^2\left(\frac{ F(t_1,x-\lambda/2)-F(t_2,x-\lambda/2)}{\lambda^2/\pi}\right) } \r)^{1/2}  \\
    &\hspace{6em}\times
   \l(  \frac{ \dow_{t_1}F\left(t_1,x+\lambda/2\right) \,
    \dow_{t_2}F\left(t_2,x+\lambda/2\right)}
    {\sinh ^2\left(\frac{ F(t_1,x+\lambda/2)
    - F(t_2,x+\lambda/2)}{\lambda^2/\pi}\right) } \r)^{1/2} \,.
}
To introduce background fields, we can express $F(t,x\pm \lambda/2)$ as a Taylor expansion around $F(t,x)$ and substitute the partial derivatives $\dow_xF$ with covariant derivatives $D_xF$. To wit
\es{}{
    F(t,x\pm \lambda/2)
    &\to \sum_{n} \frac{(\pm \lambda/2)^n}{n!}D_x^n F(t,x)\,, \\
    \p_t F(t,x\pm \lambda/2)
    &\to \sum_{n} \frac{(\pm \lambda/2)^n}{n!}
    D_tD_x^n F(t,x)\,.
}
We further need to update the measure
\begin{align}
    dt_1 dt_2 \to dt_1 dt_2\,e_t(t_1,x)e_t(t_2,x)\,.
\end{align}
Since covariant derivatives $D_t$ and $D_x$ do not commute, we have specify an ordering prescription. We chose to put time derivatives to the left of space derivatives, i.e.~$D_t^m D_x^n\bullet$.

Implementing this strategy for the non-local term results in \textit{zero} contribution to ${\cal E}^t$ and non-zero to ${\cal E}^x$.\footnote{This property only holds for the ordering of derivatives that we picked. For other orderings, we would get generate improvement terms in the energy current.} The reason for this is that the non-local term is time reparametrisation invariant, so $e_t$ disappears except in $D_x F$, where it is multiplied by $e_x$. Taking a derivative with respect to $e_t$ leaves an overall factor of $e_x$ which then is set to zero. It was previously guessed in~\cite{Abbasi:2021fcz} that energy density receives contribution from Schwarzian only, here we can derive this statement. Additionally, we can derive non-Gaussian quantum corrections absent in~\cite{Abbasi:2021fcz}.

\subsection{Classical conserved current}
\label{sec:energy-density-classical}

Recall from Section~\ref{sec:energy-diff} that the classical solution of the noise equation of motion is given by $Q=0$. Therefore, to proceed gradually, we begin by setting $Q=0$. In this limit, we find that the energy current to subleading order in derivatives is given as 
\es{EnergyCurr}{
{\cal E}^\mu= \coup N{2\pi^3\ov 3}\begin{pmatrix}
{1\ov  2D}\, (\p_{t} F)^2\\[4pt]
-\p_{t} F\p_{tx} F-{\lam^2\ov 12}\,\le(3\p_{tx} F \p_{txx} F+\p_{t} F\p_{txxx} F\ri)
\end{pmatrix}+O(\lam^4)\,.
}

Expressing $\p_t F$ in terms of ${\cal E}^t$, the constitutive relations for the energy flux take a particularly elegant form:
\es{EnergyFlux}{
{\cal E}^x=-D\le(\p_x\, {\cal E}^t+{\lam^2\ov 12}\p_{x}^3\,{\cal E}^t
+   {\lam^4\ov 360}\,\p_{x}^5\,{\cal E}^t
+O(\lam^6)\ri)\,.
}
Here we have also included the next order term in $\lambda$ that has been worked out in the supplementary Mathematica notebook.
It is straightforward to check that the continuity equation
\beq
\pr_t\, {\cal E}^t + \pr_x\, {\cal E}^x = 0
\eeq
is equivalent to the equation of motion derived from the SK effective action including higher derivative corrections given in~\eqref{EnergyFlux}. The resulting energy diffusion equation 
\es{DiffeqHighOrder}{
\pr_t\, {\cal E}^t= D\le(\p_{x}^2\, {\cal E}^t+{\lam^2\ov 12}\,\p_{x}^4\,{\cal E}^t+{\lam^4\ov 360}\,\p_{x}^6\,{\cal E}^t+O(\lam^6)\ri)
}
remains entirely linear and involves only spatial derivatives even after including higher derivative corrections. 

We can argue this feature based on  dimensional analysis. Note that the interaction term in the microscopic action~\eqref{eqn:SYK chain hamiltonian} does not have any internal dimensionful scales besides the lattice spacing $a$ (which we have set to 1). Therefore, corrections to the constitutive relations~\eqref{EnergyFlux} must take the form $\cE^x = -D\big( 1 + \hat\cO(a\p_x) \big)\dow_x\cE^t$, where $\hat\cO(a\dow_x)$ is some dimensionless operator. Here we have used the fact that $D$ is just an overall prefactor that comes from the microscopic interaction strength $\coup$ and that the interactions do not affect the definition of the energy density $\cE^t$, which is given by just the Schwarzian.
In particular, the constitutive relations cannot involve any time derivatives as the interaction term in the microscopic theory does not involve any internal time scales.
Further assuming that the constitutive relations are well-behaved and have a regular $\cE^t\to 0$ limit,\footnote{Lifting this assumption, the constitutive relations could admit corrections such as $\cE^x \sim (\dow_x\cE^t \dow_x^2\cE^t)/\cE^t$, which are singular in the limit $\cE^t\to 0$.} the operator $\hat\cO$ cannot depend on $\cE^t$ itself and thus the constitutive relations must be linear in $\cE^t$.

In fact, we can determine the non-perturbative completion of the diffusion equation. In Appendix~\ref{app:kernel_shift}, we establish that the dispersion relation of the diffusion pole is $\om= - 4iD \sin^2(\lam k/2)/\lam^2$. Combining this information from the linearity argued above, it follows that 
 the higher derivative corrections to energy diffusion in our model resum to
\es{exactDiffEq}{
\pr_t\, {\cal E}^t
&= \frac{4D}{\lambda^2} \sinh^2\left(\frac{\lambda\p_x}{2}\right) {\cal E}^t \\
\implies 
\pr_t\, {\cal E}^t(t,x)
&= D\,
\left(\frac{{\cal E}^t(t,x+\lambda)- 2{\cal E}^t(t,x)+{\cal E}^t(t,x-\lambda)}{\lambda^2} \right) \,.
}
 Note the $\lambda$ expansion of~\eqref{exactDiffEq} matches~\eqref{DiffeqHighOrder} up to $ O(\lambda^4)$.
The term in the parenthesis in the second line is just the discrete second derivative of energy density on the SYK chain.  It would be interesting to also derive this equation from a microscopic starting point in the SYK chain.

Diffusive systems satisfy the Einstein relation
\beq\label{eq:Einstein}
D = \frac{\kappa}{c_V}\,,
\eeq
where the thermal conductivity $\kappa$ and heat capacity $c_V$ are defined via
\beq
\Ec^x = -\kappa(T) \,\pr_x T\,,  \qquad 
c_V(T) = \frac{\pr \Ec^t}{\pr T}\,.
\eeq
Taking the derivatives explicitly gives
\es{kappaCv}{
\kappa(T) = gN \frac{2 \pi^3}{3}\, T\,,\qquad 
c_V(T) = {4 \pi^2 \alpha_S\, N\ov\Jc}\,T\,,
}
and we indeed recover the diffusion constant in eq.~\eqref{eq:diffusion_c}. 

\subsection{Quantum corrections and non-locality}

As we explained in Section~\ref{sec:coupling-action-to-background}, the energy flux comes from the non-local part of the action. In the derivative expansion the fluxes ${\cal E}^x_\pm$ become a local function of time, but retain non-locality in terms of SK reparametrisation variables $F^\pm$. Let us explain this point in more detail.

The leading order in derivatives expression for the ${\cal E}^x_+$ flux, as obtained in the supplementary Mathematica notebook, is given as
\beq
{\cal E}_+^x 
= -\coup N{2\pi^3\ov 3} \Big( T \pr_{x} T 
- 2iT^2 \pr_x Q + 2i Q T\pr_{x}T +  O(\lam^2) \Big)\,.
\eeq
The $\cE^x_-$ flux is obtained similarly by replacing $Q$ with $-Q$ above. The energy density does not receive any $Q$-dependent corrections at this order in $\lambda$. Recall that $T=\dow_tF$ and $F$ is the classical variable $F = (F^+ + F^-)/2=F^+-\lam^2\,Q/2$. Then we can rewrite the flux as
\es{EpwithQ}{
{\cal E}_+^x 
= -\coup N{2\pi^3\ov 3} \Big( \p_t F^+ \pr_{xt} F^+
- 2i(\p_tF^+)^2 \pr_x Q 
+ 2i Q \p_t F^+\pr_{xt}F^+ +  O(\lam^2) \Big)\,.
}
The purely $F^+$ term is localised on the forward fold, but the terms involving $Q$ also involve the backward fold, hence 
inserting ${\cal E}_+^x$ into the SK contour results in an effectively bilocal insertion into both fold. This is illustrated by the following equation: 
\beq
\includegraphics[scale=0.65]{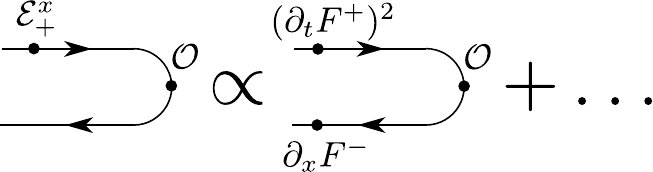}
\eeq
where the term we wrote down explicitly is taken from the second term in~\eqref{EpwithQ}.
Note that the insertions happen at the same Lorentzian time. The intuitive reason for this is the same as we gave around~\eqref{sinh}: the non-locality of the inter-site term is limited to $t\sim\beta$ in Lorentzian time (because of the $\sinh^{-1}(t/\beta)$ dependence), but does not penalise jumping to different legs of the SK contour. Contrast this to the energy density 
\es{EtSch}{
    {\cal E}^t_+ = - \frac{\lambda^4 N \alpha_S}{\Jc}
    \Sch\le[\tanh(\pi F^+/\lambda^2),t\ri]\,,
}
which is localised on the forward fold.

Because of this property, the four-point out-of-time-ordered correlation (OTOC) function of ${\cal E}^x$ can couple to shockwaves on the OTOC contour leading to exponential growth with maximal Lyapunov exponent. In contrast, ${\cal E}^t$ is a purely local expression for which we do not expect such behaviour. We leave the OTOC computation in our model for future work~\cite{wip_with_akash}.

\subsection{Local action with background fields}

In the discussion above, we coupled the non-local effective action of the SYK chain to background fields. We varied it with respect to the background fields to obtain the energy currents on the individual Lorentzian folds and afterwards implemented the derivative expansion on them.
There is an alternative avenue that we could take, wherein we directly derive the local SK effective action coupled to background fields starting from the non-local one and use it to directly read off the energy currents in their derivative expanded form.

This latter prescription requires us to introduce covariant derivatives directly on the $F$ and $Q$ fields. 
To this end, we define the average and difference combinations of energy current and background fields as\footnote{The decomposition of the background fields is a convention. For example, another convenient choice is to take $1/e^\pm_{t} = e_{t}^{-1}\pm \lam^2 e^{-1}_{q,t}/ 2$. This would simplify the definition of covariant derivatives $D_tF$ and $D_tQ$.
However, the resulting average/difference background fields do not have the nice transformation properties under diffeomorphisms as in~\eqref{decomposed-background-transformation}.}
\es{avdiffEnergy}{
    {\cal E}^\mu_\pm = {\cal E}^\mu 
    \pm \frac{\lambda^2}{2} {\cal E}_q^\mu\,, \qquad 
    e^\pm_\mu = e_\mu
    \pm \frac{\lambda^2}{2} e_{q,\mu}\,.
}
Under infinitesimal average time diffeomorphisms, the background fields $e_\mu$ and $e_{q,\mu}$ transform in the same way as~\eqref{eq:background-field-transformation}. However, under infinitesimal difference time diffeomorphisms, we find the transformations
\es{decomposed-background-transformation}{
    e_\mu
    &\to 
    e_\mu + \frac{\lambda^2}{4} 
    \left( e_{q,t} \pr_\mu \xi_q
    + \xi_q\partial_t e_{q,\mu} \right)
    + O(\xi_q^2)\,, \\
    e_{q,\mu}
    &\to e_{q,\mu} + e_t \pr_\mu \xi_q
    + \xi_q\partial_t e_{\mu}
    + O(\xi_q^2)\,.
}
Note that the flat configuration of background fields is given as $e_t = 1,\,e_x=e_{q,\mu}=0$. We observe that around a flat background configuration, $e_{q,\mu}$ transforms as a background gauge field for difference time diffeomorphisms generated by $\xi_q$. Due to the relative sign in the definition of energy current in~\cref{eq:energy-current}, one may check that 
\begin{align}
    {\cal E}^\mu = 
    - \lambda\frac{\delta I}{\delta e_{q,\mu}}, \qquad 
    {\cal E}_q^\mu = 
    - \lambda\frac{\delta I}{\delta e_{\mu}}\,.
\end{align}
This suggests that the average energy current is a Noether current of difference time translations, and vice-versa; we discuss the Noether procedure in Section~\ref{sec:Noether}. Note that both these currents are conserved on a flat background.

The above construction enables us to define the covariant derivatives of $F$ and $Q$ fields. To wit
\es{CovDerEx}{
D_t F&\equiv \frac{D_{t} F^++D_{t}F^-}{2} 
= \frac{1}{e_t}\p_t F + O(\lambda^2)\,,\\
D_t Q&\equiv \frac{D_{t} F^+-D_{t}F^-}{\lambda^2}
= \frac{1}{e_t}\p_t Q - \frac{e_{q,t}}{e_t^2}\,\p_t F+ O(\lambda^2)\,,\\
D_x F&\equiv \frac{D_{x} F^++D_{x}F^-}{2} 
= \p_x F - \frac{e_x}{e_t} \p_t F + O(\lambda^2)\,,\\
D_x Q&\equiv \frac{D_{x} F^+-D_{x}F^-}{\lambda^2}
= \p_x Q  - \frac{e_x}{e_t} \p_t Q - \frac{e_{q,x} e_t - e_x e_{q,t}}{e_t^2} \p_t F+ O(\lambda^2)\,,
}
and similarly for higher covariant derivatives. With these in place, we find that the SK effective action coupled to background fields at leading order in $\lambda$ is given as
\es{LeffCurved}{
I_\SK
&= 
\coup N{2\pi^3\ov 3\lambda}\int dt dx\,e_{t}
\Bigg[
{1\ov D} \le( D_t F \,D_tQ + \half (D_tF)^{2} \frac{e_{q,t}}{e_t} \ri)
+{Q \,(D_{t} D_x F)^2\ov D_t F}-D_xQ \,D_{t}D_xF
\\
&\hspace{16em} 
+i{\le(D_x Q\, D_t F-Q \,D_{t}D_xF\ri)^2\ov D_t F}\Bigg]+O(\lam^2)\,.
}
We observe that this effective action, except one term proportional to $e_{q,t}$, can be obtained from the effective action~\eqref{Leff2} in the absence of background fields by replacing ordinary derivatives with covariant derivatives while using the same ordering prescription, $D_t^m D_x^n\bullet$, for the (non-commuting) covariant derivatives as in the microscopic description. It would be interesting to understand the covariantisation of the effective action to higher orders in the derivative expansion.

 If we vary this action with respect to the background fields, we obtain the same results as in the previous subsections. See the Mathematica file for these checks.

\subsection{Relation to stochastic energy diffusion}\label{sec:stochastic}

Having identified the energy current, we can connect  our theory with the path integral formulation of stochastic energy diffusion. To this end, let us perform a change of variables from $F$ and $Q$ to\footnote{This field transformation is non-singular because all dependence on $F$ in the theory arises via $\p_t F$. Furthermore, the Jacobian of this transformation is ${\pi^3\coup N\ov 3\lambda D}\,\p_t$. As we discuss in Section~\ref{sec:SL2R}, the zero modes of this operator, time-independent functions $f(y)$, are not integrated over in the path integral (they are gauged). Hence the Jacobian  is simply an overall normalisation factor in the path integral.} 
\begin{align}
    {\cal E}^t
    = {\pi^3\coup N\ov 3 D}\,(\p_t F)^2
    + O(\lambda^2)
    \,, \qquad 
    \chi_a = -\frac{Q}{\p_t F}+ O(\lambda^2)\,.
\end{align}
We will formally introduce $\chi_a$ in Section~\ref{sec:avg-diff-translations} as the Goldstone of spontaneously broken difference time translations and obtain the following rewriting of the action in Section~\ref{sec:SKaction_bottomup}. In terms of these variables, the effective action takes the form
\es{SchemForm}{
{I}_\text{SK}
&= \frac{1}{\lambda}\int dt dx
\Bigg[
    \chi_a \le( \p_t{\cal E}^t + \p_{x}{\cal E}^x
\ri)
+ iT^2\kappa\,
(\dow_x \chi_a)^2
+O(\lam^2)
\Bigg]\,,
} 
where the thermal conductivity $\kappa$ is given in~\eqref{kappaCv}. The quadratic in $\chi_a$ terms above play the role of noise and $T^2\kappa$ is understood as a function of the energy density ${\cal E}^t$.
Following the case of the Brownian particle reviewed in Appendix~\ref{app:brownian}, we integrate in a new field $\xi$ through the relation
\es{xiIntIn}{
\int D\xi \,\exp\le( \frac{i}{\lambda} \int dt dx\, \le[ {i\ov2}\xi^2 
- \left(2T^2\kappa\right)^{1/2}\xi \p_x \chi_a\ri]\ri)
= \exp\le(\frac{i}{\lambda}\int dt dx\, iT^2\kappa\, (\p_x \chi_a)^2 \ri)\,.
}
Using the LHS of this equation, $\chi_a$ is now a merely Lagrange multiplier, and after integrating it out we obtain the path integral representation of correlation functions\footnote{Technically, our theory would only be equivalent to stochastic energy diffusion if $\kappa\sim 1/T^2$. In this case, the stochastic diffusion equation would be linear in $\cE^t$ and integrating it out in~\eqref{xiIntIn2} would yield
\begin{align}
    \langle\bullet\rangle = \int D\xi \,
    (\bullet)_\text{on-shell}
    \exp \left( -\frac{1}{2\lambda} \int dt dx\,\xi^2 \right)\,.
\end{align}
Here $(\bullet)_\text{on-shell}$ is evaluated on the solution of $\p_t{\cal E}^t 
+ \p_{x}( {\cal E}^x + \left(2T^2\kappa\right)^{1/2}\xi)$. However, we have $\kappa \sim T$ in our theory, so performing the $\cE^t$ integral results in a non-trivial Jacobian in the path integral. This falls in the general category of Langevin-type stochastic differential equations (SDEs) with \emph{multiplicative} noise. The aforementioned Jacobian factor drops out in the It\^o prescription of discretising time in an SDE; see e.g.~\cite{2010PhRvE..81e1113A}.}
\es{xiIntIn2}{
\langle \bullet \rangle
&= \int D F DQ\,(\bullet)\, {\rm e}^{i I_{\text{SK}}} \\
&=\int D{\cal E}^t D\xi \,
    (\bullet)\,
    \de\bigg[
    \p_t{\cal E}^t 
    + \p_{x} \Big( {\cal E}^x
    + \left(2T^2\kappa\right)^{1/2}\xi \Big)
    \bigg] 
    \exp\left(-\frac{1}{2\lambda} \int dt dx\,\xi^2\right)\,,
}
valid at leading order in $\lambda$.
This results in a stochastic energy diffusion equation (imposed by a functional delta function at every point in spacetime) with white noise
\es{WhiteNoise}{
\langle\xi(t,x) \xi(t',x')\rangle
= \lambda\,\de(t-t')\de(x-x')\,.
}
The field $\xi$ plays the role of stochastic noise contribution to the energy flux.

\section{Continuous symmetries and Noether currents}
\label{sec:Noether}

In this section we discuss various continuous symmetries of our model and the ensuing Noether currents. We will see that the energy currents in Section~\ref{sec:energy} are Noether currents corresponding to average and difference time translations on the Schwinger--Keldysh contour. A curiosity is that one of them is realised as an internal symmetry, while the other as a spacetime symmetry. Hence the average energy current will arise as a Noether current for a global shift symmetry, while the difference energy current as the time component of the energy-momentum tensor. 

The space component of the energy-momentum tensor is also conserved, but vanishes in the classical limit $Q=0$, similarly to the difference energy current; we do not know what role it plays in the dynamics of the theory. Finally, the variables that we are working with have an $\SL(2,\mathbb{R})^{2M}$ gauge redundancy, which is a property of the microscopic Schwarzian chain action~\eqref{eq:final_action} we started from. This gauge redundancy is different than the gauge symmetries of gauge theories in that it is global in time but local in space, hence the Noether method produces a nonzero gauge current for it. Another important finding will be that only one generator of the $\SL(2,\mathbb{R})$ is compatible with the derivative expansion.  

\subsection{Noether currents for higher derivative theories}

The effective theory contains higher derivatives, hence we need a higher derivative Noether formula to compute the conserved currents. Consider a field theory with dynamical fields $\phi^A$. For an infinitesimal symmetry transformation, $\phi^A\to \phi^A + \de \phi^A$, the Lagrangian may change by a total derivative
\es{Kdef}{
\de \mathcal{L}=\p_\mu K^\mu\,.
}
Then the associated Noether current is
\es{NoetherCurr}{
j^\mu=-K^\mu+\sum_A\sum_{n=0}^\infty \sum_{k=0}^n (-1)^{k}\,
\p_{a_1\dots a_k}\le({\p \mathcal{L}\ov \p\le(\p_{\mu a_1\dots a_n}\phi^A\ri)}\ri)\p_{a_{k+1}\dots a_n}\de\phi^A\,.
}
The energy-momentum tensor of a manifestly space and time translation invariant Lagrangian is accordingly given as
\es{StressTens}{
\tau^\mu_{\,\,\,\nu}=\sum_A\sum_{n=0}^\infty \sum_{k=0}^n (-1)^{k}\,
\p_{a_1\dots a_k}\le({\p \mathcal{L}\ov \p\le(\p_{\mu a_1\dots a_n}\phi^A\ri)}\ri)
\p_{a_{k+1}\dots a_n \nu}\phi^A-\de^\mu_\nu \mathcal{L}\,.
}

There is one subtlety in applying these formulas: they treat the derivatives $\p_{\mu\nu\dots}$ and $\p_{\nu\mu\dots}$ as distinct objects, and we have to compensate this by combinatorial factors.\footnote{This situation may be familiar from general relativity, where we have to take variational derivatives with respect to a symmetric tensor field.} For the special case of $1+1$ dimensions, it is also useful to rewrite
\es{jmurewrite}{
j^\mu&=-K^\mu+\sum_A\sum_{m,n=0}^\infty \sum_{k=0}^m\sum_{\ell=0}^n w_{\mu,k,\ell,m,n}\,(-1)^{k+\ell}\,
\p_{t}^k \p_x^\ell\le({\p \mathcal{L}\ov \p\le(\p_\mu\p_{t}^{m}\p_x^{n}\phi^A\ri)}\ri)
\p_{t}^{m-k}\p_{x}^{n-\ell}\de\phi^A\,,\\
w_{\mu,k,\ell,m,n}&\equiv{{k+\ell\choose k}\,{m+n-k-\ell\choose m-k}\ov {m+n+1\choose m+\de_{\mu, t}}}\,,
}
where the numerator of the weight $w_{\mu,k,\ell,m,n}$ counts the number of $\p_{a_1\dots a_{k+\ell}}$ and $\p_{a_{k+\ell+1}\dots a_{m+n}}$ combinations that are identical, while the denominator counts the identical $\p_{\mu a_1\dots a_{m+n}}$ combinations.\footnote{The weights can be fractional and are symmetric under the exchange of $t,x$ despite appearances.} These formulas are implemented in the accompanying Mathematica file for the symmetries discussed below.

\subsection{Average and difference time translations}
\label{sec:avg-diff-translations}

Let us begin with the time translation symmetry of our theory. The original time translation symmetry of the SYK chain gets doubled on the SK closed-time contour, acting independently on the forward and backward Lorentzian folds. Correspondingly, the SK EFT features two independently conserved energy currents as discussed in Section~\ref{sec:energy}. As discussed there, the average of these currents is interpreted as the classical energy current, while the difference as its stochastic/quantum noise partner.

Our microscopic SYK chain model~\eqref{eqn:SYK chain hamiltonian} is invariant under time translations. When the theory is put on the SK closed-time contour, this results in independent time translation symmetries on the forward and backward folds respectively, i.e.
\begin{align}
    t_\pm \to t_\pm - \xi_\pm,
\end{align}
where $\xi_\pm$ are constant parameters and $t_\pm \to t_\pm/\lambda^2$ are understood as rescaled time coordinates. The symmetries act on the reparametrisation fields as $F^\pm(t,x) \to F^\pm(t + \xi_\pm,x)$. It is useful to decompose the two symmetries into average and difference combinations $\xi_\pm = \xi \pm \lambda^2\xi_q/2$. In terms of the average and difference fields $F(t,x)$ and $Q(t,x)$, the average time translations simply act as overall time translations
\begin{align}\label{eq:avg-time-translation}
    F(t,x) \to F(t+\xi,x)\,, \qquad 
    Q(t,x) \to Q(t+\xi,x)\,.
\end{align}
The action of the difference time translations is quite non-trivial in this basis, but simplifies in the infinitesimal limit as
\es{eq:difference-time-translation}{
    F(t,x) 
    &\to F(t,x)
    + \frac{\lambda^4\xi_q }{4} \dow_t Q(t,x) 
    + O(\xi_q^2)\,, \\
    Q(t,x) 
    &\to
    Q(t,x)
    + \xi_q \dow_tF(t,x)
    + O(\xi_q^2)\,.
}
 Note that these infinitesimal transformations rules are valid non-perturbatively in $\lambda$. One may explicitly check that the symmetries~\eqref{eq:avg-time-translation} and~\eqref{eq:difference-time-translation} are respected by our effective action~\eqref{Leff2} at leading order in $\lambda$, while the checks at subleading orders in $\lambda$ have been performed in the supplementary Mathematica notebook.

As it turns out, the Noether current associated with difference time translations is the average energy current across the two Lorentzian folds. We have verified this by plugging in the transformation rule~\eqref{eq:difference-time-translation} into~\eqref{jmurewrite} and found that the corresponding Noether current $j^\mu$ is equal to ${\cal E}^\mu$ from~\eqref{avdiffEnergy}.
In the classical limit, i.e.~when the noise field $Q$ is set to zero, this reproduces the standard energy conservation equation.

We observe that at leading order in $\lambda$, the combination $Q/\dow_t F$ transforms as a Goldstone field, i.e.~with a simple shift, under difference time translations:
\begin{align}\label{eq:basic-goldstone}
    \frac{Q}{\dow_t F}
    &\to
    \frac{Q}{\dow_t F}
    + \xi_q 
    - \frac{\lambda^4\xi_q }{4} \frac{Q\dow_t^2 Q}{(\dow_t F)^2}
    + O(\xi_q^2)\,.
\end{align}
We can infer just from this that the difference time translations are spontaneously broken in our theory. This is a general feature of SK EFT for hydrodynamics, known as strong-to-weak spontaneous symmetry breaking~\cite{Buca:2012zz,Ogunnaike:2023qyh,Akyuz:2023lsm}. Herein, the difference combination of a global symmetry of the theory on the closed-time contour is spontaneously broken while the average one remains intact. 

A more intuitive way to demonstrate this symmetry breaking in our theory is to consider the thermal saddle point configuration $F^\pm={t\ov\beta}$, which under an average time translation acquires a shift $F^\pm\to {t\ov\beta}-{\xi\ov \beta}$. As discussed below in Section~\ref{sec:SL2R}, space-dependent but time-independent shifts are gauged in our theory, hence a compensating gauge transformation can cancel this shift and show that the thermal state is invariant under average time translations. This is not true for the difference time translation, which is broken by the thermal state.

In fact, we can correct the combination in~\eqref{eq:basic-goldstone} order-by-order in $\lambda$ to arrive at the nonlinear Goldstone field that appears in the SK EFT framework:
\es{chiaDef}{
    \chi_a = -\frac{Q}{\dow_t F} 
    - \frac{\lambda^4}{8} \frac{Q^2}{(\dow_tF)^2}
    \left( \frac{\dow_t^2 Q}{\dow_tF} 
    - \frac{Q\dow_t^3F}{3(\dow_tF)^2}
    \right)
    + { O}(\lambda^8)\,,
}
such that
\begin{align}
    \chi_a \to \chi_a - \xi_q.
\end{align}
We will discuss the connection of our $F$ and $Q$ variables with those in the conventional SK EFT framework in detail in Section~\ref{sec:bottomup}.

\subsection{Energy-momentum tensor}

The Noether current associated with average time translations is the time component of the energy-momentum tensor $\tau^\mu_{\,\,\,t}$.  We have verified by explicit computation in the derivative expansion that 
\es{}{
\tau^\mu_{\,\,\,t}= {\cal E}_q^\mu+\text{(improvement term)}\,,
}
where we use the notation from~\eqref{avdiffEnergy} for the difference energy current between the two folds.
${\cal E}_q^\mu$ identically vanishes in the classical limit, nevertheless it is very important in the computation of correlation functions, e.g.~the retarded correlator of energy currents is given by $G^{R,\mu\nu} = i \langle {\cal E}^\mu {\cal E}_q^\nu\rangle$. 

We have also computed $\tau^\mu_{\,\,\,x}$ and verified that it is conserved $\p_\mu\tau^\mu_{\,\,\,x}=0$; see the Mathematica file. It identically vanishes in the classical limit similarly to $\cE^\mu_q$. This had to be the case, as classical diffusion does not have conserved momentum. We are not aware of any use of this conserved current in the computation of correlation functions. It would be interesting to understand how it constrains the dynamics of noise in the system.

Note that it is a peculiar feature of our formalism that the canonical energy-momentum tensor $\tau^\mu_{\,\,\,\nu}$ does not capture the classical energy current ${\cal E}^\mu$ of the system, instead it contains two quantum currents.

\subsection{Reparametrisation and \texorpdfstring{$\SL(2,\mathbb{R})$}{SL(2,R)} symmetry}
\label{sec:SL2R}

A single SYK dot possesses an emergent $\SL(2,\mathbb{R})$ symmetry at low energy~\cite{Maldacena:2016upp}, which is to be treated as a gauge symmetry for the $f_x$ fields, because it leaves the fermion two-point function invariant. This gauge redundancy is different than the gauge symmetries of familiar gauge theories in that it is global in time. In our case, $\SL(2,\mathbb{R})$ can act independently on each dot. However, it must act the same way on the forward and backward folds of the SK contour, because the non-local part of the action couples them. In the continuum limit, the infinitesimal action of $\SL(2,\mathbb{R})$ on the reparametrisation fields $F^\pm$ read as
\beq
    F^{\color{red}\pm} \ra F^{\color{red}\pm} + \delta_{\boldsymbol 0}(x) 
    + \delta_{\boldsymbol +}(x)\, e^{2\pi F^{\color{red}\pm}/\lambda^2} 
    + \delta_{\boldsymbol -}(x)\, e^{-2\pi F^{\color{red}\pm}/\lambda^2}
    \,, 
\eeq
where $(\delta_{\boldsymbol 0}, \delta_{\boldsymbol +}, \delta_{\boldsymbol -})$ are independent parameters of the transformation on each lattice site.\footnote{Note that these ${\boldsymbol +}/{\boldsymbol -}$ (in bold) are $SL(2,\mathbb{R})$ indices rather than SK contour indices. We again used the red font for the latter to emphasise that.} 

One of the transformations, $\delta_{\boldsymbol 0}$, simply shifts the classical variable $F=(F^{+}+F^{-})/2$ by a constant and leaves the quantum variable $Q=\frac{F^+ - F^-}{\lambda^2}$ invariant. This symmetry is obvious in the hydrodynamic expansion, since $F$ only enters via $\pr_t F$. 

However, the other two components of the $\SL(2,\mathbb{R})$ symmetry are not manifest in the hydrodynamic expansion. 
For example, $\delta_{\boldsymbol +}$ acts as 
\es{plusTf}{
\delta_{\boldsymbol +}: \quad & F \ra F +  \delta_{\boldsymbol +} 
\,e^{2\pi F/\lambda^2} \cosh\le(\pi\,Q\ri)\,,\\
& Q \ra Q + {2\delta_{\boldsymbol +}\ov \lam^2} \,e^{2\pi F/\lambda^2} \sinh\le(\pi\,Q\ri)\,.
}
These expressions have non-uniform $\lam$ scaling and hence spoil the grading. In particular, the shifts are exponentially large in the $\lam$ expansion. Around a fixed time, $t=t_0$, we could make the shift of $F$ to be $O(\lambda^0)$ by rescaling as $\delta_{\boldsymbol +}(x)=\widetilde\delta_{\boldsymbol +}(x)\, e^{-2\pi F(t_0,x)/\lam^2}$, however the shift of $Q$ would still be $O(\lambda^{-2})$ at this time. Also, taking time derivatives of both sides shows another violation of grading, as the $\pr^n_t F$ shifts by $O(\lambda^{-2n})$ from the term when all derivatives act on the exponential.\footnote{In terms of unscaled time, the $n$th time derivative should scale as $\lambda^{2n}$, but its shift would be $O(\lambda^0)$.}
The same problem occurs for the $\delta_{\boldsymbol -}$ generator:
\es{minusTf}{
\delta_{\boldsymbol -}: \quad & F \ra F +  \delta_{\boldsymbol -} 
\,e^{-2\pi F/\lambda^2} \cosh\le(\pi\,Q\ri)\,,\\
& Q \ra Q - {2\delta_{\boldsymbol -}\ov \lam^2} \,e^{-2\pi F/\lambda^2} \sinh\le(\pi\,Q\ri)\,.
}

We conclude that the $\delta_{\boldsymbol \pm}$ symmetries cannot be manifest in the derivative expansion. It would be interesting to investigate whether truncating the action to a finite order in the derivative expansion leads to any pathology because of the broken gauge symmetry. We also note that $\SL(2,\mathbb{R})$ transformations (on a fourfold contour) are important in describing scrambling~\cite{Maldacena:2016upp,Blake:2017ris,Blake:2021wqj}.

The $\SL(2,\mathbb{R})$ action is spatially local, but global in time, hence there is a corresponding Noether current. Since it is only the $\delta_{\boldsymbol 0}(x)$ transformation that is consistent with the derivative expansion, we focus on the current associated to it. The current vanishes in the classical limit, for $Q=0$. We did not find its explicit formula enlightening even to $O(\lam^0)$, hence we relegate  it to Appendix~\ref{app:quantumNoether}. In a schematic form the current takes the form
\es{SL2schematic}{
{\cal G}^\mu &= \coup N{2\pi^3\ov 3}\begin{pmatrix}
\delta_{\boldsymbol 0}\,\bullet+(\p_x \delta_{\boldsymbol 0})\,\bullet\\[4pt]
\delta_{\boldsymbol 0}\,\bullet
\end{pmatrix}+O(\lam^2)\,,\\
\p_\mu {\cal G}^\mu&=-\delta_{\boldsymbol 0}(x)\,{\de I_{\text{SK}}\ov \de F(t,x)}=0\,,
}
where $\bullet$'s stand for  expressions containing at least one $Q$, and in the second line we found that the conservation equation of the current is proportional to the quantum equation of motion. Note that we get infinitely many conserved currents, as ${\cal G}^\mu$ is conserved for any choice of $\delta_{\boldsymbol 0}(x)$. This is not a surprise, since in the continuum limit the symmetry is local, $\SL(2,\mathbb{R})^\infty$.

\section{Dynamical KMS symmetry}
\label{sec:dynKMS}

In this section we discuss the discrete KMS (Kubo-Martin-Schwinger) symmetry of our SK EFT. This symmetry emerges as the macroscopic incarnation of the microscopic time reversal symmetry. We further use the KMS symmetry to derive the entropy current for our theory.

The original SYK chain model is time reversal symmetric on the microscopic level. However, the effective action expressed in terms of the reparametrisation field is clearly not time reversal invariant. The mechanism for this violation is the same as in the classic Caldeira-Leggett~\cite{kamenev2023field} model: in order to arrive at~\eqref{eq:final_action} on the SK contour, we effectively integrated out a continuum of states at arbitrary low energies, leading to dissipation. Mathematically it is reflected in $i \epsilon$ making $\sinh^2$ a non-even function and hence the action not being invariant under:
\beq
\label{eq:T_conj}
\Theta: F^{\pm}(t) \ra -F^{\pm}(-t)\,.
\eeq
The Schwarzian part of the action is still invariant under this transformation. However, for the inter-site piece, the time reversal transformation~\eqref{eq:T_conj} gets updated to the dynamical KMS symmetry transformation of the low-energy effective action.

\subsection{KMS symmetry of thermal correlation functions}

KMS symmetry is an elementary, but very consequential property of thermal correlators. By inserting operators on the SK contour on Figure~\ref{fig:SK},  we get a correlator with time-ordered and anti-time-ordered insertions: 
\es{}{
\tr\bigg(\rho_\text{th} \, 
\bar{\cal T} \{A(t_1) \dots A(t_K) \}\, 
{\cal T} \{B(t_{K+1}) \dots B(t_L) \}\bigg)\,,
}
where $\rho_{\text{th}} = {\rm e}^{-\beta H}/Z$ denotes the thermal density matrix, while ${\cal T}$ and $\bar{\cal T}$ denote time-ordering and anti-time-ordering operators. $A$ and $B$ may represent any operators of interest in the theory; in particular, energy density or flux. We have suppressed the spatial dependence of the operators for simplicity. Noting that the thermal density matrix $\rho_{\text{th}}$ implements a translation in imaginary time, $\rho_\text{th} A(t)=A(t+i\tau)\rho_\text{th}$, and using cyclicity of trace, the same correlator can also be written as
\es{KMSStatement_derivation}{
& \tr\bigg(\rho_\text{th} \, 
\bar{\cal T} \{A(t_1) \dots A(t_K) \}\, 
{\cal T} \{B(t_{K+1}) \dots B(t_L) \}\bigg) \\
&= \tr\bigg(\rho_\text{th} \, 
{\cal T} \{B(t_{K+1}) \dots B(t_L) \}\,
\bar{\cal T} \{A(t_1+i\beta) \dots A(t_K+i\beta) \}
\bigg) \\ 
&=\tr\bigg(\rho_\text{th}\,
{\cal T} \{B(t_{K+1}-i\beta/2) \dots B(t_L-i\beta/2) \}\,
\bar{\cal T} \{A(t_1+i\beta/2) \dots A(t_K+i\beta/2) \}
\bigg)\,,
}
where in the third line we performed an imaginary time translation of all operators by $i\beta/2$. To constitute a symmetry, we need to bring the correlator to the original form, where time-ordered operators are inserted to the right of anti-time-ordered operators, which is not the case in the last line above. To bring it to the desired form, we perform a time reversal transformation. Since time-reversal is an anti-unitary transformation, it reverses the ordering of operator insertions. To wit,
\es{KMSStatement}{
\eqref{KMSStatement_derivation}&=\eta_A^K\eta_B^{L-K}\tr\bigg(\rho_\text{th} \,
\bar{\cal T} \{A(-t_K - i\beta/2) \dots A(-t_1- i\beta/2) \} \\
 &\hspace{10em}
{\cal T} \{B(-t_{L}+i\beta/2) \dots B(-t_{K+1}+i\beta/2) \}
\bigg)\,,
}
Here $\eta_{A,B}$ are the eigenvalues of $A,B$ under time reversal, e.g.~$\Theta A(t)=\eta_A A(-t)$.\footnote{Time reversal is an anti-unitary symmetry and acts on correlation functions for complex time arguments as $\tr(\rho_{\text{th}} A(z_1)B(z_2)) = \tr(\rho_{\text{th}} A(-z^*_1)B(-z^*_2))^* = \tr(\rho_{\text{th}}B(-z_2)A(-z_1))$.} This is known as the KMS symmetry of thermal correlation functions.

\subsection{Dynamical KMS symmetry}

The KMS symmetry~\eqref{KMSStatement} above assumed exact thermal equilibrium with Gibbs density matrix $\rho_{\text{th}}=e^{-\beta H}/Z$.
However, in real life, things are rarely in exact thermal equilibrium and there generically exists a small temperature gradient. Therefore, it was conjectured in~\cite{Crossley:2015evo,Glorioso:2016gsa,Glorioso:2017fpd} that the all-order hydrodynamic action must posses a \textit{dynamical} KMS symmetry, which reproduces the KMS symmetry~\eqref{KMSStatement} for correlation functions in thermal equilibrium. 

In Appendix~\ref{app:KMS}, we demonstrate that
the effective action~\eqref{eq:final_action} is in fact invariant under the following discrete symmetry transformation expressed in terms of the continuum reparametrisation fields
\begin{subequations}
\es{KMSCont-pm}{
&F^\pm(t,x) \ra -F^\pm(-t,x) \pm \frac{i\lambda^2}{2} \,,
}
or equivalently in average-difference basis
\es{KMSCont}{
&F(t,x) \ra -F(-t,x)\,, \\
&Q(t,x) \ra -Q(-t,x)+i\,.
}\label{eq:KMS-cont-together}%
\end{subequations}
Note that this symmetry implements a shift by $i$ on the noise field $Q$, thereby supporting our choice of  $\lambda^0$ grading for $Q$ in~\eqref{FieldCM}.
In fact, collecting all terms with a fixed number of derivatives yields an expression which is a finite polynomial in $Q$. However, note that this symmetry violates the natural $N$ scaling discussed around~\eqref{Leff4}. We further comment on this feature in Appendix~\ref{app:brownian}, where we compare with the case of the Brownian particle.

The transformations~\eqref{eq:KMS-cont-together} do not constitute the true dynamical KMS symmetry of the theory, as they do not directly reproduce the KMS symmetry~\eqref{KMSStatement} for operator insertions on the forward and backward folds of the closed-time contour. However, recall from Section~\ref{sec:avg-diff-translations} that our theory also has a difference time translation symmetry. We can combine it with~\eqref{eq:KMS-cont-together} to read off the true KMS symmetry transformation
\begin{subequations}
\es{simpleshifts}{
&F^\pm(t,x) \ra -F^\pm\le(-t \pm i\lambda^2\beta/2,x\ri) \pm \frac{i\lambda^2}{2}\,.
}
Note that this symmetry leaves the classical saddle point configuration $F^\pm(t,x) = t/\beta$ invariant. Note also that $\beta$ is a constant unrelated to the dynamical temperature $T=\p_t F$ which may emerge under the effect of sources.

Since difference time translations act quite non-trivially on $F$ and $Q$ fields, the action of the KMS symmetry on these variables is also quite non-trivial. However, it simplifies at leading order in $\lambda$, i.e.
\es{fulldynKMS}{
F(t,x) 
&\ra 
- \cos\left(\frac{\lambda^2\beta}{2}\p_t\right) F\le(-t,x\ri) 
- \frac{i\lambda^2}{2} \sin\left(\frac{\lambda^2\beta}{2}\p_t\right) Q\le(-t,x\ri) \\
&\ra 
- F\le(-t,x\ri) +  O(\lambda^4)\,, \\
Q(t,x) 
&\ra 
- \cos\left(\frac{\lambda^2\beta}{2}\p_t\right) Q\le(-t,x\ri) 
- \frac{2i}{\lambda^2}\sin\left(\frac{\lambda^2\beta}{2}\p_t\right) F\le(-t,x\ri)
+ i \\
&\ra 
- Q\le(-t,x\ri) 
- i \Big( \beta\p_tF\le(-t,x\ri) - 1\Big) +  O(\lambda^4)
\,,
}\label{eq:dynKMS}%
\end{subequations}
where the trigonometric functions of derivatives implement the shifts in~\eqref{simpleshifts}.
Note that unlike some $\SL(2,{\mathbb R})$ transformations in Section~\ref{sec:SL2R}, the action of the KMS transformation is homogeneous in $\lambda$ and thus respects the derivative expansion. Note that, just like~\eqref{KMSCont}, the dynamical KMS transformation does not respect the natural $N$ scaling of $Q$ discussed around~\eqref{Leff4}.\footnote{Further comments are given on this issue in Appendix~\ref{app:brownian}, around~\eqref{QKMS2}.}

It is an interesting open question to write down the most general KMS symmetric effective action. Our theory is special in that all terms in the Lagrangian are of the form
\es{SchematicForm}{
(Q-i/2)^k \p_t^l \,\p_x^m\,\le[ Q^n \, F^p\ri]\,, \qquad k+l+n+p=\text{even}\,,
}
where the derivatives are distributed on the $Q$'s and $F$'s in the square bracket in such a way that at least one derivative acts on each field. We have verified that this property of the SYK chain hydrodynamic effective action follows order-by-order from the steps presented in Section~\ref{sec:loc_from_nonloc}. Such terms are automatically invariant under the discrete transformations~\eqref{eq:KMS-cont-together} without the need of dropping any total derivatives. However, enforcing that a combination of such terms obey the difference time translation symmetry~\eqref{eq:difference-time-translation} and thus the dynamical KMS symmetry~\eqref{eq:dynKMS} is considerably more involved and transforms the Lagrangian by total derivatives.

\subsection{Fluctuation-dissipation theorem}
\label{sec:fluctdiss}

The relation between fluctuations and dissipation is a cornerstone result of statistical physics. It follows from the general structure of the SK EFT. Nevertheless, it is interesting to spell out its various manifestations in our concrete system.

Let us consider the system in thermal equilibrium and linearise around the saddle as in~\eqref{SaddleLin}.
The fluctuation-dissipation theorem (FDT) relates the two point functions $\langle f q\rangle$ (dissipation) and $\langle ff\rangle$ (fluctuation). In Fourier-space, it states
\es{FDT}{
\Big\langle f(\omega,k)f^*(\omega,k) \Big\rangle 
= {\lam^2}\coth\le({\lam^2\beta\om\ov 2}\ri)
\Re\Big\langle f(\omega,k) q^*(\omega,k) \Big\rangle\,.
}
 Note that $f^*(\omega,k)=f(-\omega,-k)$ and similarly for $q$.
This follows by a straightforward application of the KMS symmetry~\eqref{KMSStatement} to two-point functions with different choices of contour. In terms of rescaled time, the length of the Euclidean circle is $\lam^2 \beta$, which results in the factor of $\lambda^2$ in the argument of $\coth$ above. Furthermore, the fields on the $\pm$ folds of the closed-time contour are given as $f\pm \lam^2 q/2$, giving the overall $\lam^2$ factor in~\eqref{FDT}.

The linearised SK effective action in Fourier-space takes the schematic form
\es{LinAct}{
I_{\SK,2} 
= \half \int_{-\infty}^\infty \frac{d\omega \,dk}{(2\pi)^2} 
\begin{pmatrix}
{f^*(\om,k)} && {q^*(\om,k)}
\end{pmatrix}
\begin{pmatrix}
0 &&  A(\om, k)\\
 {A^*(\om,k)} &&  B(\om, k)
\end{pmatrix}
\begin{pmatrix}
{f(\om, k)} \\ {q(\om, k)}
\end{pmatrix}\,,
}
for some coefficients $A(\omega,k)$ and $B(\omega,k)$.
Note that the $11$ entry of the coefficient matrix is zero because the effective action contains no terms involving only $F$. 
From here we can obtain the correlation functions
by inverting the kinetic matrix: 
\es{twopointfns}{
\Big\langle f(\omega,k)f^*(\omega,k) \Big\rangle 
= \frac{-iB(\omega,k)}{|A(\omega,k)|^2}, \qquad 
\Big\langle f(\omega,k)q^*(\omega,k) \Big\rangle 
= \frac{i}{A^*(\omega,k)}\,.
}
Imposing the FDT, we find that $A$ and $B$ must be related as
\es{FDTABrel}{
B(\omega,k)
= -i\lam^2 \coth\le({\lam^2\beta\om\ov 2}\ri)
\Im A(\omega,k)\,.
}
We have verified that this relation holds up to $O(\lam^4)$ in the action.\footnote{For completeness, we record the expressions:
\es{ABexpr}{
A(\om, k)
&= \frac{4\pi^2 \alpha_SN}{\lambda\Jc}\le[\om^2+{\lam^4 \beta^2\omega^4\ov 4\pi^2}\ri]
-i\omega {2\pi^3 gN \over 3\lambda}\le[k^2-{\lam^2k^4\ov 12}+\lam^4\le({k^6\ov 360}+{1\ov 4 \pi^2}k^2(\beta\omega)^2\ri) \ri]
+O(\lam^6) \,,\\
B(\om, k)&=i{4\pi^3 gN\over 3\lambda\beta}\le[k^2-{\lam^2k^4\ov 12}+\lam^4\le({k^6\ov 360}+{(\pi^2+3)\ov 12 \pi^2}k^2(\beta\omega)^2\ri)\ri]
+O(\lam^6)\,.
}
Note that the real term in $A$ is the Schwarzian contribution and it does not contribute to~\eqref{FDTABrel}.
} We also verify to $O(\lam^4)$ that the retarded correlation function $G^R\equiv i\langle f q\rangle$ has a diffusion pole with dispersion relation $ \om= - 4iD \sin^2(\lam k/2)/\lam^2$, in complete agreement with the linearised results reviewed in Appendix~\ref{app:kernel_shift}.

The relation~\eqref{ABexpr} also follows from the dynamical KMS symmetry transformation~\eqref{eq:dynKMS}. Note that in Fourier space,~\eqref{eq:dynKMS} implies
\es{}{
\begin{pmatrix}
    f(\omega,k) \\
    q(\omega,k)  
\end{pmatrix}
= -
\begin{pmatrix}
    \cosh\left(\frac{\lambda^2\beta\omega}{2}\right)
    & -\frac{\lambda^2}{2} \sinh\left(\frac{\lambda^2\beta\omega}{2}\right) \\
    -\frac{2}{\lambda^2}\sinh\left(\frac{\lambda^2\beta\omega}{2}\right)
    & \cosh\left(\frac{\lambda^2\beta\omega}{2}\right)
\end{pmatrix}
\begin{pmatrix}
    f(-\omega,k) \\
    q(-\omega,k)  
\end{pmatrix}\,.
}
Enforcing that this transformation leaves the linearised effective action~\eqref{LinAct} invariant, we precisely recover~\eqref{FDTABrel}.

\subsection{Entropy current}
\label{sec:entropy}

An elegant consequence of the dynamical KMS symmetry is that it leads to an entropy current $\cS^\mu$ whose divergence is locally non-negative. This was proven in~\cite{Glorioso:2016gsa} in full generality in the SK EFT framework by using only the unitary properties of the SK effective action together with KMS symmetry. 
Here we use a slightly different algorithm presented in~\cite{Armas:2024iuy} to extract the entropy current in our theory. 

Two Lagrangians give rise to the same action if they differ by a total derivative. Our SK EFT has multiple symmetries. We can always make the Lagrangian exactly invariant under one symmetry, but under another symmetry transformation the Lagrangian may change by a total derivative as in~\eqref{Kdef}. For the construction of the entropy current $\Sc^\mu$, we consider the SK Lagrangian $\cL_\SK(F,Q)$ in the form~\eqref{Leff2}, such that it is exactly invariant under the symmetry transformation~\eqref{eq:KMS-cont-together}, without the need of total derivative terms. This transformation composes the KMS transformation with a difference time translation.

We know that the Lagrangian must take the form $\cL_\SK(F,Q) = -\hat\cL_{(1)}Q + O(Q^2)$, where $\hat\cL_{(1)}$ is some differential operator acting on $Q$. The operator $\hat\cL_{(1)}$ has information regarding the classical equations of motion. In fact, 
\begin{align}\label{eq:L1-expansion}
    \text{EoM} = \lambda \frac{\delta I_\SK}{\delta Q}\bigg|_{Q=0}
    \quad\implies\quad 
    - \lambda\hat\cL_{(1)}Q = Q\,(\text{EoM}) 
    + \dow_\mu\big( \hat\cS^\mu Q \big)\,,
\end{align}
where we have extracted the total derivative term left over from variational derivatives for some linear differential operator $\hat\cS^\mu$ acting on $Q$. The entropy current is given as $\cS^\mu = \hat\cS^\mu 1$; provided that the SK Lagrangian is exactly invariant under~\eqref{eq:KMS-cont-together}. Note that the entropy current is a classical object, which has not been generalised to include noise terms.

Following~\cite{Glorioso:2016gsa}, the local entropy production rate is given as
\begin{subequations}\label{eq:second-law}
\begin{align}\label{eq:second-law-statement}
    \dow_\mu \cS^\mu = -(\text{EoM}) + \Delta,
\end{align}
where
\begin{align}\label{eq:simple-Delta}
    \Delta = \lambda\int_{-\infty}^\infty dq\,
    \frac{\pi\cosh(\pi q)}{\sinh^2(\pi q)}
    \Im\cL_\SK(F,q)\,.
\end{align}
\end{subequations}
Technically,~\eqref{eq:second-law} only applies to theories where $\Im\cL_\SK$ is even under time reversal, which is true for our case.
We have presented a self-contained general proof of the entropy production rate in Appendix~\ref{app:secondlaw}. 

We can use the algorithm above to read off the entropy current and entropy production rate for our theory. To leading order in $\lambda$, we find
\es{EntropyCurr}{
\Sc^\mu=\coup N{2\pi^3\ov 3}\begin{pmatrix}
{1\ov D}\, T\\[4pt]
- \p_{x} T
\end{pmatrix}+O(\lam^2)\,,\qquad
\Delta = \coup N{2\pi^3\ov 3} {\le(\p_{x}T\ri)^2\ov T}
    +  O(\lambda^2)\,.
}
Note that the entropy density satisfies the first law of thermodynamics
\begin{align}
    d{\cal E}^t = T d{\cal S}^t\,,
\end{align}
understood to apply to leading order in derivatives.
One may verify that the entropy current satisfies the second law~\eqref{eq:second-law} upon using the equation of motion~\eqref{DiffEq}.
We have included the entropy current calculation to $O(\lam^4)$ in our theory in the supplementary Mathematica notebook and have explicitly checked that~\eqref{eq:second-law} holds to that order.

After using the equations of motion, the total on-shell entropy production is given as
\begin{align}\label{eq:total-entropy}
    S_{\text{tot}} = \int dtdx\, \Delta 
    = \lambda\int_{-\infty}^\infty dq\,
    \frac{\pi\cosh(\pi q)}{\sinh^2(\pi q)}\Im I_\SK[F,q]\,,
\end{align}
One of the unitarity conditions in SK EFTs is that $\Im I_\SK\geq 0$ for all field configurations. Since it is multiplied by a positive factor, it follows that $S_{\text{tot}}\geq 0$. However, hydrodynamics is built around the notion of local thermodynamic equilibrium. Therefore, we expect a stronger local version of the second law to hold in hydrodynamics, i.e.~$\Delta \geq 0$ at every point in spacetime. This would require that $\Im\cL_\SK \geq 0$ locally, which is not guaranteed by $\Im I_\SK\geq 0$. Note that this is true in our theory at leading order in $\lambda$, as can be seen in~\eqref{EntropyCurr}, but a more general understanding is desirable.

Fortunately, the algorithm outlined above to extract the entropy current leaves room for improvements. In particular, we can improve the SK Lagrangian with total derivative terms that respect~\eqref{eq:KMS-cont-together}, for instance $\p_t^2 \left(Q(1+i Q)/\p_tF\right)$, which accordingly induces improvements to $\cS^\mu$ and $\Delta$. We prove that these improvements can be tuned in such a way that $\Im\cL_\SK \geq 0$ to any order in $\lambda$.\footnote{What we mean by this is the following: for a generic fluid configuration, this condition is satisfied by the leading order expression of $\Delta$ at every spacetime point. However, it may happen that at a spacetime point $(t_0,x_0)$ the entropy production rate $\Delta$ vanishes to $O(\lam^k)$, e.g. if~$\p_x T(t_0,x_0)=0$ in our theory (see~\eqref{EntropyCurr}). Then, the all order non-negativity of $\Delta$ means that the $\lam^{k+1}$ term is required to be positive.} In our theory, we find that the improved $\Im\cL_\SK$ can be expressed as a complete square
\begin{align}
    \Im\cL = 
    {2\pi^3 \coup N \ov 3\lambda T}
    \bigg( T \p_x Q  -Q \p_{x}T 
    + O(\lam^2)
    \bigg)^2
\end{align}
This guarantees via~\eqref{eq:simple-Delta} that $\Delta\geq 0$ locally also to a given order in $\lambda$. See the details of the construction in Appendix~\ref{app:secondlaw} and the supplementary Mathematica notebook.\footnote{Note that the $\Delta$ generated through this procedure, while guaranteed to be non-negative, is not generically expressed as a simple complete square itself. This should be contrasted with the construction of~\cite{Glorioso:2016gsa}.} 

\subsection{Stochastic transport coefficients and higher powers of $Q$}

As discussed in~\cite{Jain:2020zhu}, the dynamical KMS symmetry only uniquely determines the effective action of an SK EFT in terms of the classical constitutive relations up to cubic order in the fields. As explained before, the $O(Q)$ part of the effective action precisely characterises the classical constitutive relations.
Hence the results of~\cite{Jain:2020zhu} imply $O(Q^2)$ and $O(Q^3)$ parts of the action are completely characterised by the $O(Q)$ part up to cubic orders in interactions.

Let us illustrate this at $O(Q^2)$. There is a dissipation-fluctuation relation in the EFT that applies far from equilibrium. We focus on the leading order in the derivative expansion. Some $ O(Q)$ terms in the Lagrangian give rise to dissipation in the equations of motion, while the $O(Q^2)$ terms control the fluctuations; these get related by the dynamical KMS symmetry. Under time reversal the action~\eqref{Leff2} transforms as\footnote{Note that the time reversed Lagrangian $\cL_{\SK}(\Theta F,\Theta Q)$ is evaluated at $-t$, which we have flipped back to $t$ in the expression above. Since the action is obtained by integrating the Lagrangian over all space and time, this distinction does not matter and merely corresponds to a change of integration variables.}
\beq
\l( gN \frac{2\pi^3}{3\lambda} \r)^{-1} 
\bigg(\cL_{\SK}(\Theta F,\Theta Q)\Big|_{(-t)\to t} - \cL_{\SK}(F,Q) \bigg) 
= 2\p_x Q \pr_{tx} F - 2Q \frac{(\pr_{tx}F)^2}{\pr_t F} + \Oc(\lambda^2)\,.
\eeq
To constitute a symmetry, the extra linear terms in $Q$ must be cancelled by shifting the $ O(Q^2)$ terms according to $Q\to Q+i$~\eqref{KMSCont}:
\es{KMSshiftFDT}{
&\left(\coup N{2\pi^3\ov 3\lambda} \right)^{-1}
\bigg( \cL_\SK (F,Q+i) - \cL_\SK(F,Q) \bigg) \\
&= \underbrace{{i(\p_{tx} F)^2\ov \p_t F}}_{\text{from shifting} \  O(Q)}
- \underbrace{i{\le(\p_{tx}F\ri)^2\ov \p_t F}
+ 2{(\p_x Q \p_t F-Q \p_{tx}F)\p_{tx}F \ov \p_t F}}_{\text{from shifting} \  O(Q^2)}
\,+\, O(\lam^2)\,.
}
We see that at the order $ O(Q^0)$, the cancellation of the term $(\pr_{tx} F)^2/\pr_t F$ requires that the coefficients in front of $ O(Q)$ and $ O(Q^2)$ terms in the original Lagrangian conspire. The same is true for cancelling the $ O(Q)$ contribution. A more complicated version of this arguments carries over to $ O(Q^3)$~\cite{Jain:2020zhu}.

 However, quartic or higher terms like $(\dow_{tx}F)^2 (\p_xQ)^2$, $(\dow_{tx}F) (\p_xQ)^3$, or the higher-noise sector $O(Q^{n})|_{n\geq 4}$, may admit new \emph{stochastic transport coefficients} that are independent of the classical constitutive relations. These are understood as parametrising independent microscopic information about the system that does not affect the classical dynamics of hydrodynamic observables and is only visible through fluctuations.

In our theory, we find that one such stochastic transport coefficient $\vartheta_2(T)$ is present at $O(\lam^2)$. To wit, we have 
\es{eq:st-transport}{
I_\SK
&\supset \coup N{2\pi^3\ov 3\lambda} \int dtdx 
\frac{-7i\pi^2}{60(\partial_t F)^3} 
\left[
\lambda^2\Big( Q\partial_{tx}F - \dow_tF\partial_x Q\Big)^2
\Big((Q-i)\partial_{tx}F-\dow_tF\partial_x Q\Big)^2
+ O(\lam^6)
\right]\\
&\supset \frac{1}{\lambda} \int dtdx\,i\vartheta_2(T)
\bigg[
\lambda^2\left(\partial_x \chi_a\right)^2
\left(T^2\partial_x\chi_a - i\partial_{x}T\right)^2
+ O(\lam^6)
\bigg]\,
}
where we have used the definition of $\chi_a$~\eqref{chiaDef}, and we have identified
\begin{align}
    \vartheta_2(T) 
    = -\frac{7\pi^2}{60}\kappa(T) = 
    -\coup N{2\pi^3\ov 3}\le(
    \frac{7\pi^2}{60}\ri) T\,.
\end{align}
The complete effective action at $O(\lam^2)$ is given in Appendix~\ref{app:highorder}.
One may check that the quartic contribution in~\eqref{eq:st-transport} is invariant under the dynamical KMS symmetry up to $O(\lambda^6)$ for arbitrary $\vartheta_2(T)$. Generically, the theory of energy diffusion can admit another such stochastic transport coefficient $\vartheta_1(T)$ at $O(\lam^2)$ identified in~\cite{Jain:2020zhu}, but it is identically absent in $1+1$-dimensional systems.\footnote{Generically, the two stochastic transport coefficients are given as
\begin{align}
    \cL_{\SK}/\lambda \supset
    \vartheta_1(T)\Big(\nabla \chi_a\cdot \left(T^2\nabla\chi_a - i\nabla T\right) \Big)^2
    + \Big(\vartheta_2(T) - \vartheta_1(T)\Big)
    (\nabla \chi_a)^2 \left(T^2\nabla\chi_a - i\nabla T\right)^2
    + \cO(\lam^4)\,.
\end{align}
}

\section{The Schwinger--Keldysh EFT for energy diffusion}
\label{sec:bottomup}

In this work we have derived an SK EFT for the SYK chain model with nearest neighbour interactions. In this section, we compare our theory to the bottom-up SK EFT formalism for energy diffusion~\cite{Crossley:2015evo, Blake:2017ris}. We find the map of our $F,Q$ variables and symmetries to those more commonly used in the literature, and fix all transport coefficients.

\subsection{General setup}

\paragraph*{Fields and global symmetries}

The fundamental ingredients of the SK EFT for energy diffusion are a pair of time coordinate fields $X_{\pm}(\sigma,x)$, defined on a fluid worldvolume (worldsheet in $1+1$ dimensions) with internal time parameter $\sigma$ and space coordinate $x$. This is similar to the Lagrange description of a fluid, but concerning the time coordinate instead of space.\footnote{Recall that the Euler description of the fluid is given in terms of fluid velocity field $u^x(t,x)$. By contrast, the Lagrange description is in terms of a coordinate field $X^x(t,y)$ such that $u^x(t,X^x(t,y)) = \dow_t X^x(t,y)$. Similarly in our case, the Euler description of energy diffusion is given in terms of the temperature field $T(t,x)$, which is expressed in the Lagrange-like description as $T(X(\sigma,x),x)^{-1} = \beta\dow_\sigma X(\sigma,x)$, where $\beta$ is a reference inverse temperature. } We can think of the $X_+$ and $X_-$ fields as living on the forward and backward folds of the closed-time contour respectively. The theory features a pair of global time translation symmetries that act independently on the $X_+$ and $X_-$ fields, i.e.
\begin{align}\label{eq:SK-time-translation}
    X_\pm(\sigma,x) \to X_\pm(\sigma,x) - \xi_\pm.
\end{align}
We can introduce background fields for these symmetries, $e^\pm_\mu$ from Section~\ref{sec:energy}. The respective covariant derivatives are given as
\es{}{
    D_\sigma X_\pm(\sigma,x) &= e^\pm_t(X_\pm(\sigma,x),x)\, \dow_\sigma X_\pm(\sigma,x)\,, \\
    D_x X_\pm(\sigma,x) &= e^\pm_t(X_\pm(\sigma,x),x)\, \dow_x X_\pm(\sigma,x)
    + e^\pm_x(X_\pm(\sigma,x),x)\,.
}
(Note that $\sigma$ here is not a spacetime index.) These objects are invariant (not just covariant) under time diffeomorphisms
\begin{align}\label{eq:SK-time-diffeo}
    X_\pm(\sigma,x) \to X_\pm(\sigma,x) - \xi_\pm\big(X_\pm(\sigma,x),x\big)\,.
\end{align}
higher derivatives, like $\dow_\sigma D_\sigma X_\pm$, are automatically invariant under time diffeomorphisms. 

\paragraph*{Worldvolume reparametrisation symmetry}

The theory is required to be invariant under \emph{space-dependent} time reparametrisations on the worldvolume
\begin{align}\label{eq:WV-repar}
    \sigma \to \sigma - a(x)\,,
\end{align}
which acts on the fields as: $X_\pm(\sigma,x) \to X_\pm(\sigma-a(x),x)$. Physically, this symmetry amounts to a freedom of arbitrarily relabelling the internal clock on the worldvolume without affecting the physical observables. The theory is also invariant under global spatial translation symmetries (and rotation symmetries in higher spatial dimensions) on the worldvolume, which act on the fields as usual.

\paragraph*{Unitarity conditions}

The SK effective action $I_\SK[X_+,X_-]$ is generally complex. However, requiring that it emerges from a unitary microscopic theory results in three non-trivial unitarity conditions~\cite{Liu:2018kfw}:
\begin{align}
\label{eq:unitarity-general}
    I_\SK[X,X] =0\,, \qquad 
    I_\SK[X_-,X_+] = -I^*_\SK[X_+,X_-]\,, \qquad 
    \Im I_\SK[X_+,X_-] \geq 0\,.
\end{align}
Broadly speaking, they arise from fundamental properties of density matrices in quantum mechanics: $\tr\rho =1$, $\rho^*=\rho$, and $0\leq\text{Eigenvalues}(\rho)\leq 1$ respectively. The unitarity conditions naturally generalise in the presence of background fields, with $X_\pm$ replaced with $\{X_\pm,e^\pm_\mu\}$.

\paragraph*{KMS symmetry}

Lastly, to describe correlation functions in the thermal state, the SK EFT needs to be invariant under a dynamical KMS symmetry, which acts on the dynamical and background fields as
\es{eq:KMS-SK}{
    X_\pm(\sigma,x) 
    &\to -X_\pm\big({-}\sigma \pm i\lambda^2\beta/2,x\big) \pm i\lambda^2\beta/2\,, \\
    e^\pm_t(t,x)
    &\to e^\pm_t(-t\pm i\lambda^2\beta/2,x)\,, \\
    e^\pm_x(t,x)
    &\to -e^\pm_x(-t\pm i\lambda^2\beta/2,x)\,.
}
The factors of $\lambda$ appear above because we are working in rescaled coordinates.
Note that the dynamical KMS symmetry leaves the equilibrium state $X_\pm(\sigma,x) = \sigma$ invariant. It also leaves any time independent background field configurations invariant, up to an overall time-reversal flip. Similarly to~\Cref{sec:dynKMS}, we can define a KMS shift symmetry by combining the KMS symmetry with a constant difference time translation. This acts as simply
\es{}{
    X_\pm(\sigma,x) 
    &\to -X_\pm\big({-}\sigma \pm i\lambda^2\beta/2,x\big)\,, \\
    e^\pm_t(t,x)
    &\to e^\pm_t(-t,x)\,, \\
    e^\pm_x(t,x)
    &\to -e^\pm_x(-t,x)\,.
}

The dynamical KMS symmetry acts differently on the forward and backward folds of the Lorentzian contour, making it quite hard to implement during bottom-up constructions of SK EFTs. Generally, one is only able to implement the KMS symmetry in the \emph{classical regime}, where the characteristic frequencies of fluctuations lie sufficiently below the quantum scale, i.e.~$\lambda^2\omega \ll 1/\beta$.\footnote{Restoring the reduced Planck's constant $\hbar$, this condition is $\lambda^2\omega \ll 1/(\hbar\beta)$.} We define the average/difference basis of fields $X_\pm = X \pm \lambda^2 X_a/2$, on which the KMS shift symmetry on $X_\pm$ simplifies in this regime:
\es{eq:classical-KMS}{
    X(\sigma,x) 
    &\to - X(-\sigma,x) 
    + O(\lambda^4\beta^2)\,, \\
    X_a(\sigma,x) 
    &\to - \left( X_a
    + i\beta \dow_\sigma X
    \right)(-\sigma,x) 
    + \beta\,O(\lambda^4\beta^2)\,,
}
which is easier to ensure in bottom-up constructions of SK EFTs. Note that, for the continuum hydrodynamic description to be valid, we have already assumed that $\lambda^2\omega \ll D/a^2$, where $a$ denotes the microscopic lattice spacing and recall that $D$ is the energy diffusion constant. Consequently, SK EFTs that only respect the classical KMS symmetry~\eqref{eq:classical-KMS} are valid assuming the hierarchy of scales $\lambda^2\omega \ll D/a^2 \ll 1/\beta$.

By contrast, in this work we have derived the SK EFT for energy diffusion starting from the microscopic model of an SYK chain. Therefore, our theory respects the full quantum KMS symmetry, order-by-order in $\lambda$, with the validity regime $\lambda^2\omega \ll D/a^2 \sim 1/\beta$. 

\subsection{Mapping between degrees of freedom}

We can map between the $X_\pm$ variables of the SK formalism to our $F$ and $Q$ variables. Since $F$ and $Q$ variables are defined for real time instead of worldvolume time, we first define the $F^\pm(t,x)$ fields as point-wise inverses of the $X_\pm(\sigma,x)$ fields,~\footnote{The $X_\pm(\sigma,x)$ fields are monotonic in $\sigma$, hence they are invertible.} i.e.
\begin{align}
    X_\pm\Big(\beta\,F^\pm(t,x),x\Big) = t \,.
\end{align}
Having done that, the $F$ and $Q$ variables are simply given via $F^\pm = F + \lambda^2 Q/2$.
One may check that with these definitions, the action of time translations, space-dependent reparametrisations, and dynamical KMS transformations take the form as found in previous sections. In particular, note that the space-dependent reparametrisations map to the $\delta_{\boldsymbol 0}(x)$ component of the $\SL(2,{\mathbb R})$ symmetry described in Section~\ref{sec:SL2R}.
It is also useful to note that
\begin{align}
    D_\sigma X_\pm = \frac{1}{\beta\,D_t F^\pm}\,, \qquad 
    D_x X_\pm = - \frac{D_xF^\pm}{D_tF^\pm}\,,
\end{align}
where the left-hand sides are evaluated at $\sigma = \beta F^\pm$. The unitarity conditions~\eqref{eq:unitarity-general} map in these variables to
\begin{align}
\label{eq:unitarity}
    I_\SK[F,Q=0] =0\,, \qquad 
    I_\SK[F,-Q] = -I^*_\SK[F,Q]\,, \qquad 
    \Im I_\SK[F,Q] \geq 0\,.
\end{align}

Let us also take this opportunity to map to a different choice of real time fields $\sigma(t,x)$ and $\chi_a(t,x)$ that are typically employed in the literature; see e.g.~\cite{Blake:2017ris}. These are given as
\begin{align}
    X(\sigma(t,x),x) = t\,, \qquad 
    \chi_a(t,x) = X_a(\sigma(t,x),x)\,.
\end{align}
These fields have nice global time translation and reparametrisation properties:
\begin{equation}
\begin{alignedat}{3}
\text{Average time translations:}&\qquad &
\sigma(t,x) &\to \sigma(t + \xi, x)\,, \qquad &
\chi_a(t,x) &\to \chi_a(t + \xi, x)\,, \\
\text{Difference time translations:}&\qquad &
\sigma(t,x) &\to \sigma(t, x)\,, \qquad &
\chi_a(t,x) &\to \chi_a(t, x) - \xi_q, \\
\text{Reparametrisations:}&\qquad &
\sigma(t,x) &\to \sigma(t, x) - a(x)\,, \qquad &
\chi_a(t,x) &\to \chi_a(t, x)\,.
\end{alignedat}
\end{equation}
This should be contrasted with the rather complicated action of difference time translations on $F$ and $Q$. In particular, we note that the difference time translations are spontaneously broken in the SK EFT framework, with $\chi_a$ acting as the associated Goldstone. However, the action of KMS transformations is quite complicated in this basis. 
The transformation only simplifies in the classical limit
\es{}{
    \sigma(t,x) &\to -\sigma(-t,x) +  O(\lambda^4\beta^2)\,, \\
    \chi_a(t,x) &\to - \l( \chi_a(-t,x) + \frac{i\beta}{\p_t \sigma(t,x)} - i\beta \r)
    +  O(\lambda^4\beta^2)\,.
}
The KMS shift symmetry acts in the same way but with the constant $-i\beta$ part of the transformation dropped out. 

The exact map between $\sigma$, $\chi_a$ variables and our $F$, $Q$ variables is quite complicated. However one can find these order-by-order in $\lambda$, i.e.
\es{}{
    \frac{1}{\beta}\sigma
    &= F - \frac{\lambda^4}{4}
    \frac{Q}{\dow_t F}
    \left(\dow_t Q - \frac{Q\dow_t^2F}{2 \dow_tF}\right)
    +  O(\lambda^8)\,, \\
    \chi_a 
    &= -\frac{Q}{\dow_t F} 
    - \frac{\lambda^4}{8} \frac{Q^2}{(\dow_tF)^2}
    \left( \frac{\dow_t^2 Q}{\dow_tF} 
    - \frac{Q\dow_t^3F}{3(\dow_tF)^2}
    \right)
    + { O}(\lambda^8)\,.
}
We found the same $\chi_a$ empirically in~\Cref{sec:Noether} in~\eqref{chiaDef} while constructing the Goldstone field for strong-to-weak spontaneous symmetry breaking.

\subsection{Schwinger--Keldysh effective action}\label{sec:SKaction_bottomup}

Let us write down the bottom-up SK effective action for energy diffusion at leading order in $\lambda$. In terms of the average-difference combinations $X(\sigma,x)$, $X_a(\sigma,x)$, the effective action takes the form~\cite{Liu:2018kfw, Blake:2017ris, Armas:2024iuy}
\begin{align}\label{eq:SK-action-wv}
    I_\SK &= \int d\sigma\,dx
    \Bigg[ {-}\cE^t(T) D_\sigma X_a  
    - T\kappa(T) 
    \left( D_x X_a - \frac{D_x X}{D_\sigma X} D_\sigma X_a \right)
    \left( \dow_\sigma D_x X
    - \frac{D_xX}{D_\sigma X} \dow_\sigma D_\sigma X
    \right) \nonumber\\
    &\hspace{14em}
    + \frac{i}{\beta} T\kappa(T) 
    \left( D_x X_a - \frac{D_x X}{D_\sigma X} D_\sigma X_a \right)^2 
    + O(\lam^2)
    \Bigg]\,,
\end{align}
where the covariant derivatives are defined as
\es{}{
D_\sigma X = \frac{D_\sigma X_+ + D_\sigma X_-}{2}\,, \qquad
D_\sigma X_a = \frac{D_\sigma X_+ - D_\sigma X_-}{\lam^2}\,, \\
D_x X = \frac{D_x X_+ + D_x X_-}{2}\,, \qquad
D_x X_a = \frac{D_x X_+ - D_x X_-}{\lam^2},
}
while the coefficients $\cE^t$ and $\kappa$ are taken to be functions of local temperature $T = (\beta D_\sigma X)^{-1}$. Note that this worldvolume definition of local temperature matches our previous definition $T= \dow_tF$ at leading order in $\lambda$.

The action~\eqref{eq:SK-action-wv} is clearly invariant under the average and difference time translation symmetries~\eqref{eq:SK-time-translation} and has appropriate coupling to background fields due to the use of covariant derivatives. One may explicitly check that it is also invariant under the worldvolume reparametrisation symmetry~\eqref{eq:WV-repar}. It respects the unitarity conditions~\eqref{eq:unitarity-general}, provided that we take $\kappa(T)\geq 0$. However, it is only invariant under the classical KMS symmetry~\eqref{eq:classical-KMS}. 

We can express the action in the $\sigma$, $\chi_a$ basis. At leading order in $\lambda$, we find that the SK effective action takes the form
\es{eq:action-chiabasis}{
    I_\SK 
    &=\int dt dx\,e_t\left[{-} \cE^t(T) D_t\chi_a 
    + i\kappa(T)\, D_x \chi_a  \bigg(
    T^2 D_x \chi_a
    - i\left( D_xT + TE_x \right) 
    \bigg)
    + O(\lam^2)
    \right]\,,
}
where $T = \beta^{-1}\dow_t\sigma/e_t + O(\lambda^2)$ and $E_x$ generates the power term in the energy conservation equation~\eqref{eq:conservation-sourced}.\footnote{We have obtained this action in the absence of background fields before in~\eqref{SchemForm}.} The covariant derivatives are defined as
\es{}{
    D_xT
    &= \dow_x T - \frac{e_x}{e_t} \dow_t T\,,\\
    D_t\chi_a &= \dow_t \chi_a + \frac{\chi_a}{e_t} \dow_t e_t + \frac{e_{q,t}}{e_t}\,, \\
    D_x\chi_a &= e_t \dow_x \chi_a + \chi_a \dow_t e_x + e_{q,x}
    - e_x \left( \dow_t \chi_a + \frac{\chi_a}{e_t} \dow_t e_t + \frac{e_{q,t}}{e_t}\right)\,.
}
These definitions are covariant under average time translations, and are invariant under difference time translations at leading order in $\lambda$. 
We have explicitly checked that the action~\eqref{eq:action-chiabasis} matches the one derived in this work at leading order in $\lambda$ when mapped to $F$, $Q$ variables, with the coefficients $\cE^t(T)$ and $\kappa(T)$ given in Section~\ref{sec:energy}.

We emphasise that the main virtue of the SK effective action derived in this work is that, through the chain of steps shown in Figure~\ref{fig:logic}, we uncovered how the hydrodynamical variables are embedded in the microscopic ones and we determined all the transport coefficients from microscopics.
Our SK effective action is also more general than the bottom-up construction of SK EFTs described in this section. Firstly, it systematically accounts for higher derivative hydrodynamic corrections. Furthermore, while the action~\eqref{eq:SK-action-wv} or~\eqref{eq:action-chiabasis} only respect the KMS symmetry in the classical regime~\eqref{eq:classical-KMS}, our action is compatible with the fully quantum KMS symmetry at arbitrary orders in $\lambda$. There has been some recent progress on constructing bottom-up SK EFTs compatible with quantum KMS symmetry in the context of charge diffusion~\cite{Jain:2026obh}. It will be interesting to extend these results to the case of energy diffusion and compare with the results of this work, which we leave for the future.

\section{Outlook}
\label{sec:outlook}

In this paper, we gave new arguments for the microscopic soft mode action~\eqref{eq:final_action} of the SYK lattice valid for energies much smaller than the coupling $J$.
From this starting point we obtained the nonlinear theory of fluctuating hydrodynamics~\eqref{Leff2} including higher derivative corrections. We have computed a few subleading corrections, but our formalism allows one to compute them algorithmically to arbitrary order relatively easily. We  also proved the presence of the dynamical KMS symmetry conjectured to be present in any hydrodynamic theory, and we obtained the energy and entropy currents. We obtained a perfect match with the bottom-up construction of the Schwinger--Keldysh EFT of energy diffusion, after finding the appropriate mapping of the degrees of freedom.

The explicit all-order hydrodynamic action obtained in this paper can be used to address many interesting questions; let us list some of them here:

\paragraph{Additional charges}
The only conserved quantity for the Majorana SYK lattice considered in this paper is the energy.
One obvious generalisation is the addition of extra charges, for example a U(1) charge. A single complex SYK dot with a U(1) charge has an extra ``phase-reparametrisation'' field in its low-energy effective action~\cite{Gu:2019jub}. It would be interesting to find the counterpart of this theory for the SYK lattice and subsequently derive the associated SK EFT with energy and charge diffusion.

\paragraph{Quantum-complete SK EFT}

We compared our results to the bottom-up approach to the  SK EFT for energy diffusion at leading order in $\lambda$. As we discussed, a bottom-up SK EFT construction for higher derivative energy diffusion, consistent with fully quantum dynamical KMS symmetry, is missing in the literature. In light of our results, it will be interesting to revisit this question along the lines of~\cite{Jain:2026obh}. In particular, we hope to find a closed-form non-perturbative version of our SK EFT valid at arbitrary $\lambda$, compatible with the non-perturbative energy diffusion equation~\eqref{exactDiffEq} for our model. We expect that such an SK EFT will also manifest the $\SL(2,{\mathbb R})$ symmetry of the SYK chain.

\paragraph{Stochastic transport coefficients} 

We have found a stochastic transport coefficient at $O(\lam^2)$ in our theory, which belongs to a class of free parameters in SK EFTs that are only visible through fluctuations and do not enter the classical equations of motion~\cite{Jain:2020zhu}. In general, we expect many more such coefficients to appear in our theory at higher orders in $\lambda$. Not much is known about these coefficients at present, besides that they leave signatures in correlation functions via stochastic or quantum loop effects. Our theory provides one of the few non-equilibrium models where exact analytic calculations can be performed, which we hope to leverage to test and study these stochastic transport coefficients in more detail.

\paragraph{Entanglement entropy and OTOC} The action~\eqref{eq:final_action} is valid for any time contour. In this paper we considered the standard SK contour but it would be interesting to explore multi-fold time contours relevant for OTOC~\cite{wip_with_akash} and entanglement entropy (both conventional and entanglement ``in time'' \cite{Doi:2023zaf,Milekhin:2025ycm}) which differ only in boundary conditions; the  works~\cite{Stanford:2021bhl,Mezei:2025fcd} on Brownian SYK may provide important clues. Multi-fold contours also appear in the computation of spectral-form factor and their hydrodynamics has been addressed in~\cite{Winer:2020gdp,Chen:2023hra}.
We refer to~\cite{Mishra:2025vlf} for recent progress in the direction of the OTOC. 

\paragraph{Gravity dual} A single low-temperature SYK dot has a dual gravitational description in terms of two-dimensional Jackiw--Teitelboim gravity~\cite{Maldacena:2016upp}. It is an interesting question whether~\eqref{eq:final_action} can arise from higher-dimensional gravity.

\paragraph{Quantisation} 
The large $N$ limit has allowed us to treat~\eqref{eq:final_action} semiclassically. It would be interesting to understand $1/N$ corrections which become important when the temperature is of order $1/N$. The Schwarzian theory is exactly solvable~\cite{Bagrets:2016cdf,Mertens:2017mtv,Stanford:2017thb}, with free-energy being one-loop exact, so the challenge comes from the non-local part. For two coupled SYK dots a step in this direction was taken in~\cite{Milekhin:2021cou}. It was shown that keeping only the one-loop contribution to the density of states was not enough to match finite $N$ numerical computation. This question is important for higher-dimensional black holes: the Schwarzian action can be seen as a spherical reduction of the Einstein action, whereas the non-local action comes from integrating out light matter~\cite{Milekhin:2021cou}. It was recently discovered that classical effects from light matter fields near black hole horizons can lead to large (and sometimes singular) changes in the geometry and thermodynamics~\cite{Horowitz:2022mly}. However, the fate of these effects on the quantum level at low temperatures is yet to be understood.

\paragraph{Generalisation to open quantum systems}
Another interesting generalisation which should be accessible with our formalism is the case of open system dynamics, when the SYK lattice is coupled to a bath or undergoes a measurement process. The dynamics of the SYK reparametrisations under the presence of measurements has been previously investigated in~\cite{Milekhin:2022bzx}.

\paragraph{Experimental realisation}

There are several proposals on the experimental realisation of an SYK dot~\cite{Danshita_2017,Luo:2017bno,Pikulin_2017,Chen_2018,Wei_2021,Baumgartner:2024ysk}. One such proposal~\cite{Chen_2018}, which was recently realised in a lab~\cite{Anderson_2024}, is through a graphene flake with an irregular boundary in an external magnetic field. There $0+1$ dimensional SYK fermions are realised as electrons sitting on the lowest Landau level with disorder coming from the shape of the irregular boundary. It is a natural question whether the lattice model studied in this paper can be realised using known materials.

\section*{Acknowledgments}

We thank Luca Delacr\'etaz, Alexei Kitaev, and Mukund Rangamani for useful discussions. MB is supported by the Maryam Mirzakhani Graduate Scholarship. 
AJ and MM are supported by the ERC Consolidator Grant GeoChaos-101169611.
AM acknowledges funding provided by the Simons
Foundation (Grant 376205), the DOE QuantISED program (DE-SC0018407), and the Air Force Office of Scientific Research (FA9550-19-1-0360). AM also acknowledges support from the Mendeleev fellowship during his visit to the London Institute for Mathematical~Sciences.

This work is funded by the European Union. Views and opinions expressed are however those of the authors only and do not necessarily reflect those of the European Union or the European Research Council Executive Agency. Neither the European Union nor the granting authority can be held responsible for them.

For the purpose of open access, the authors have applied a CC BY public copyright licence to any Author Accepted Manuscript (AAM) version arising from this submission.

\appendix

\section{Ladder kernel eigenvalue shift}
\label{app:kernel_shift}

The Schwinger--Dyson equations for the two--point function $G_x(\tau-\tau')= \frac{1}{N} \sum_i \bra \Psi_{i,x} (\tau) \Psi_{i,x}(\tau') \ket$ in our model read
as
\beq
\pr_\tau G_x(\tau) - \int d\tau' \  \Sigma_x(\tau-\tau') G_x(\tau') = \delta(\tau) \,, \quad \text{(no $x$ sum)}\,,
\eeq
with the self-energy
\beq
\Sigma_x(\tau) = J_0^2 G_x^{p-1}(\tau) + J_1^2 \l( (1-h) G_{x-1}^{p h} G_{x}^{p(1-h)-1}(\tau)  + h  G_{x}^{ph-1} (\tau) G_{x+1}^{p(1-h)}(\tau) \r)\,, \quad \text{(no $x$ sum)}\,.
\eeq
This self-energy corresponds to the following term in the $G- \Sigma$ action:
\beq
\Delta S = \frac{J_1^2}{2p} \int d\tau_1 d\tau_2  \sum_x G_{x}^{ph} G_{x+1}^{p(1-h)} \,.
\eeq

Equilibrium solution with homogeneous ($x$-independent) $G_x \equiv G_\text{eq}$ coincides with the standard single-dot SYK-$p$ answer with effective $J^2=J_0^2+J_1^2$.
As with a single SYK dot, the four-point function is governed by the ladder kernel $K$. The kernel can also be seen as governing the fluctuations $\delta G_x$ around the equilibrium solution $G_\text{eq}$. Explicitly, we can write it as 
\beq
K = K_\text{local} + K_\text{hopping},
\eeq
and it is convenient to define each kernel via its convolution $*$ action on the perturbation~$\delta G_x(\tau)$:
\es{Kernels}{
K_\text{local}\, \delta G_x& = (p-1) J^2 G_\text{eq} * G_\text{eq}^{p-2} \delta G_x * G_\text{eq}\,,\\
K_\text{hopping} \,\delta G_x &= p h (1-h) J_1^2 G_\text{eq} * G_\text{eq}^{p-2} (\delta G_{x+1} + \delta G_{x-1} - 2 \delta G_x) * G_\text{eq}\,.
}
In fact, $K_\text{local}$ coincides with a single SYK dot ladder kernel. Following~\cite{Gu:2016oyy}, we also observe that $K_\text{hopping}$ is proportional to the lattice Laplacian times $K_\text{local}$. It is at this point that we are using the fact that the hoping term is marginal.
In the low energy limit, $K_\text{local}$ develops a series of eigenvalues close to $1$, which correspond to time reparametrisations.
Since the actual four-point function is proportional to $1/(1-K)$, these modes become dominant.
Using the known expression for these $K_\text{local}$ eigenvalues, the eigenvalues of $K$ close to $1$ can be easily obtained:
\beq
\label{eq:k_shift}
1-K_{n,k} \approx \alpha_K \frac{|n|}{\beta \Jc} - \frac{p}{p-1} \frac{J_1^2}{J^2} h (1-h)
 2 (\cos(k)-1)\,, 
 \eeq
where $k$ is the momentum along the chain and $n$ is the discrete Matsubara frequency. To write this expression we assumed that $\frac{J_1^2}{J^2} (\cos(k)-1)$ is small. 
Obviously, the expansion in terms of $k^2$ becomes the expansion in terms of the lattice Laplacian in higher dimensions.

On the other hand, performing the expansion of
\beq
\frac{J_1^2 b^p}{8  p} \int_0^\beta d\tau_1 d\tau_2\Biggl[ \l(  \frac{ f_x'\left(\tau _1\right) f_x'\left(\tau _2\right)}{\sin ^2\left(\frac{ f_x\left(\tau _1\right)-f_x\left(\tau
   _2\right)}{2}\right) } \r)^{h}  \l( \frac{ f_{x+1}'\left(\tau _1\right)  f_{x+1}'\left(\tau _2\right) }{ \sin ^2\left(\frac{ f_{x+1}\left(\tau _1\right)-f_{x+1}\left(\tau
   _2\right)}{2}\right) } \r)^{1-h} \Biggr]
\eeq
near equilibrium: $f_x = 2\pi \tau/\beta + \epsilon_x(\tau) = 2\pi \tau/\beta + \frac{1}{2\pi}\sum_n \ep_{x}(n) e^{2 \pi i n \tau/\beta} $ we get
\beq
-\frac{J_1^2 b^p}{12 p} h (1-h) \sum_ n  |n| (n^2-1) (\ep_x(n) - \ep_{x+1}(n) )^2.
\eeq
Using the normalisation of the four-point function, namely the conversion coefficient from the action to the eigenvalue shift~\cite{Maldacena:2016hyu}
\es{ConvFact}{
\frac{6p^2}{(p-1) J^2 b^p}\,.
}
 one can check that this action exactly reproduces the eigenvalue shift~\eqref{eq:k_shift}.

The kernel eigenvalue (\ref{eq:k_shift})  essentially gives rise to the diffusion pole: the continuation to the real time is performed by $|n| \ra \pm i \omega$ (depending whether we aim for the advanced/retarded propagator):
\begin{align}
& \frac{\alpha_K}{2 \pi \Jc} \l( \pm i \omega - \frac{p}{p-1} \frac{J_1^2}{J^2} h(1-h) 2 (\cos(k) -1 ) \frac{2 \pi \Jc}{\alpha_K} \r) = \\ \nonumber 
& = \frac{\alpha_K}{2 \pi \Jc} \l( \pm i \omega -  J_1^2 h(1-h) 2 (\cos(k) -1 ) \frac{2 \pi \Jc}{\alpha_S} \frac{(p-1) b^p}{6 p} \r)\,, 
\end{align}
where we recognise the diffusion constant~\eqref{Dresc}; the $h=1/2$ special case appears in the main text in~\eqref{eq:diffusion_c}.

\section{Parity-violating hydrodynamics at $h\neq 1/2$}\label{app:hdep}

The theory has non-trivial dependence on $h$, in particular it breaks parity at $h\neq \frac12$. Nevertheless the classical equation of motion is the same as for $h=1/2$ up to $O(\lam^3)$ with the rescaling 
\es{Dresc}{
D_h=4h(1-h)\,{2\pi^3 \coup\ov 3}\Big/\le(4 \pi^2 \alpha_S\ov\Jc\ri)\,.
}
The first new term with nontrivial (i.e.~not related to the $h=1/2$ terms by the rescaling factor $4h(1-h)$) is
\es{Lh}{
&{\cal L}_\SK/\le[4h(1-h)\,\coup N{2\pi^3\ov 3}\ri]\\
&\supset\, \frac{i\pi^2(1-2h)^2\lambda^2(\partial_{tx}F)^2}{20\partial_t F}
\left(\left(\frac{(\partial_{tx}F)^2}{(\partial_t F)^2}-\frac{3\partial_{txx}F}{(\partial_t F)}\right)Q^2
- (\partial_x Q)^2\right)\,.
}
The first parity breaking term occurs at $O(\lam^3)$ and is given by
\es{Lh2}{
&{\cal L}_\SK/\le[4h(1-h)\,\coup N{2\pi^3\ov 3}\ri]\\
&\supset \frac{2i\pi^2(h-1)h(2h-1)\lambda^3(\partial_{tx}F)^3}{36(\partial_t F)^2}
\left(2\left(\frac{(\partial_{tx}F)^2}{(\partial_t F)^2}-\frac{2\partial_{txx}F}{\partial_t F}\right)Q^2 - (\partial_x Q)^2\right)\,. 
}

\section{Comparison with the Brownian particle}\label{app:brownian}

It is instructive to compare and contrast our results to the paradigmatic SK EFT of a Brownian particle. We will follow the presentation of Section~III.~C. of~\cite{Liu:2018kfw}. In the large mass limit the SK EFT to leading order in derivatives is given by
\es{Brownian}{
L=-q(M\ddot{x}+k x+\nu \dot{x})+{i\ov 2}{\sig\ov \hbar} q^2\,.
}
The dynamical KMS transformation to leading order in derivatives is
\es{BrownianKMS}{
{x}(t)&\to x(-t)\,,\\
{q}(t)&\to q(-t)-i\hbar \beta \dot{x}(-t) \,.
}
We can make it more manifest that quantum fluctuations are small, by introducing $q=\hbar \tilde{q}$, then the path integral becomes
\es{PathInt}{
Z=\int DxD\tilde q\, e^{iS/\hbar}=\int DxD\tilde q\, \exp\le[i\int dt\,\le(-\tilde q(M\ddot{x}+k x+\nu \dot{x})+{i\ov 2}\sig\tilde q^2 \ri) \ri]\,,
}
and the KMS transformation of $\tilde{q}$ does not contain $\hbar$ anymore:
\es{BrownianKMS2}{
{\tilde q}(t)&\to \tilde q(-t)-i \beta \dot{x}(-t) \,.
}
The SK EFT only has KMS symmetry, if 
\es{EinsteinRel}{
\nu=\frac12\beta\sigma\,.
}
This relation is the Einstein relation of Brownian motion. To see this more explicitly, following~\cite{Liu:2018kfw} we integrate in a Hubbard-Stratonovich field using the relation
\es{PathInt2App}{
\exp\le[i\int dt\,{i\ov 2}\sig\tilde q^2  \ri]=\int D\xi\,\exp\le[i\int dt\,\le({i\ov 2\sig}\xi^2+\xi \tilde q\ri)  \ri]\,,
}
and integrate out $\tilde q$ to obtain
\es{PathInt3}{
Z=\int DxD\xi\, \de\le(M\ddot{x}+k x+\nu \dot{x}-\xi\ri)\,\exp\le[-\int dt\,{1\ov 2\sig}\xi^2 \ri]\,.
}
This is just a Langevin equation (imposed as a functional delta function) with the Gaussian white noise
\es{WhiteNoiseApp}{
\langle \xi(t) \xi(t')\rangle=\sig \de(t-t')\,.
}
Then~\eqref{EinsteinRel} is indeed the Einstein relation connecting the noise to friction through the equipartition relation (or Boltzmann distribution for the energy of the Brownian particle).

We note in passing that the Hubbard-Stratonovich trick can be generalised to the case when we have non-Gaussian noise. We introduce $\xi$ as a Lagrange multiplier together with  another auxiliary field $\chi$ 
\es{PathInt2}{
\exp\le[i\int dt\,\left({i\ov 2}\sig\tilde q^2 +\lam \tilde q^3+\dots\right) \ri]=\int D\xi D\chi\,\exp\le[i\int dt\,\le(\xi (\tilde q-\chi)+{i\ov 2}\sig\chi^2 +\lam \chi^3+\dots\ri) \ri]\,.
}
If $\lam$ and all implicit higher order terms are zero, we can integrate out $\chi$ and obtain~\eqref{PathInt2App}. If however, these are present, we proceed as in~\eqref{PathInt3}, and integrate out $q$ to get a Langevin equation, but this time with a non-Gaussian noise distribution
\es{Pxi}{
P[\xi]=\int D\chi\,\exp\le[i\int dt\,\le(-\xi \chi+{i\ov 2}\sig\chi^2 +\lam \chi^3+\dots\ri) \ri]\,.
}

In large $N$ systems in general and in the SYK chain in particular, $1/N$ plays an analogous role to $\hbar$. Indeed, we introduced $\tilde Q$ that had $O(1)$ fluctuations. In our system the  KMS transformation~\eqref{fulldynKMS}, the analogue of~\eqref{BrownianKMS}, does not include any power of $N$, but otherwise takes a very similar form
\es{QKMS}{
Q(t,x) 
\ra 
- Q\le(-t,x\ri) 
- i \Big( \beta\p_tF\le(-t,x\ri) - 1\Big)
\,.
}
If however, we aim to use $\tilde Q$, then it transform by a large shift, mixing different orders in the $1/N$ expansion:
\es{QKMS2}{
\tilde Q(t,x) 
\ra 
- \tilde Q\le(-t,x\ri) 
- i\sqrt{N} \Big( \beta\p_tF\le(-t,x\ri) - 1\Big)\,.
}
 Nevertheless, if $F=t/\beta+O(1/\sqrt{N})$, i.e.~it is close to the global thermal state, then the natural $N$ scaling of $Q=O(1/\sqrt{N})$ is respected. Such an $F$ would be natural in the case, when we turn on infinitesimal sources to compute thermal correlators. However, if we drive the state far from equilibrium with large sources, $F$ can deviate from the thermal value by an arbitrary amount.

The dynamical KMS symmetry relates different coefficients to each other in the SK EFT, leading to fluctuation-dissipation relations and the Einstein relation discussed around eqs.~\eqref{FDT} and~\eqref{eq:Einstein}, which are the analogues of~\eqref{EinsteinRel}.

\section{Proof of the KMS shift symmetry~\eqref{eq:KMS-cont-together} }\label{app:KMS}

In this appendix we prove the KMS shift symmetry \eqref{eq:KMS-cont-together} for our theory. Let us begin with the Schwarzian part of the low-energy effective action~\eqref{eq:final_action}. One can check that it is actually invariant under a larger symmetry $F(t) \to -F(-t + ib/2) + i\theta/2$ for any constants $b$ and $\theta$. Recall from \cref{eq:sch-lorentzian-twofolds} that the respective contribution to the SK EFT $I_{\SK,\Sch}$ is simply the difference of the Schwarzian action evaluated on the two Lorentzian folds. Since $F^+$ and $F^-$ fields have no interactions in $I_{\SK,\Sch}$, it is actually invariant under a larger symmetry 
\begin{align}\label{eq:general-shift}
    F^\pm(t) \to -F^\pm (-t + ib^\pm/2) + i\theta^\pm/2,
\end{align}
for independent constants $b^\pm$ and $\theta^\pm$.

Let us now consider the inter-site coupling part of the action $I_{\SK,\text{non-loc}}$ in \eqref{Itot}. We begin with the interaction term $I_{+-}$. Setting $\lambda=1$, we have
\es{eq:interaction-app}{
    \frac{iI^\epsilon_{+-}}{gN\pi^2} 
    &=
    \int dt_1 dt_2
    \l(  \frac{ F_x'^+(t_1+ i\epsilon) F_x'^-(t_2-i\epsilon)}
    {\sinh^2\left(\pi F_x^+(t_1+ i\epsilon)-\pi F_x^-(t_2-i\epsilon) \right) } \r)^{1/2} \\
    &\hspace{6em}
    \l(  \frac{ F_{x+1}'^+(t_1+ i\epsilon) F_{x+1}'^-(t_2-i\epsilon)}
    {\sinh^2\left(\pi F_{x+1}^+(t_1+ i\epsilon)-\pi F_{x+1}^-(t_2-i\epsilon) \right) } \r)^{1/2}.
}
The factors of $\pm i\epsilon$, for some small positive $\epsilon$ implement the correct contour ordering while performing the time integrals. This effectively deforms the contour for difference time-integral $u\equiv t_1-t_2$ in imaginary time by $+i\epsilon$. Note that the numerical value of the integral does not depend on the precise magnitude of $\epsilon$, given that it obeys $\epsilon\ll \beta_0$, with $\beta_0$ the characteristic inverse temperature scale of the $F$ field configurations.\footnote{To this to hold, we assume that the $F$ functions are analytic in a strip of width $\sim\beta_0$.}  This is because the respective contour deformation does not cross any singularities arising from the sinh in the denominator. One can check that $I_{-+}$ is given by the same expression as above after performing a change of integration variables $t_1\leftrightarrow t_2$.

Note that \eqref{eq:interaction-app} is not invariant under the transformation \eqref{eq:general-shift} for arbitrary $b^\pm$ and $\theta^\pm$. In particular, it is not invariant under the trivial time-reversal transformation $\theta^\pm=b^\pm=0$ due to the $i\epsilon$ prescription in the argument of $\sinh$. However, if we choose $\theta^\pm = \pm 1$ and $b^\pm = \pm b$, we find the transformed action
\es{eq:interaction-app-transformed}{
    \frac{i\tilde I^\epsilon_{+-}}{gN\pi^2} 
    &=
    \int dt_1 dt_2
    \l(  \frac{ F_x'^+(-t_1+i\tilde\epsilon) F_x'^-(-t_2-i\tilde\epsilon)}
    {\sinh^2\left(-\pi F_x^+(-t_1+i\tilde\epsilon)+\pi F_x^-(-t_2-i\tilde\epsilon)+i\pi \right) } \r)^{1/2} \\
    &\hspace{6em}
    \l(  \frac{ F_{x+1}'^+(-t_1+i\tilde\epsilon) F_{x+1}'^-(-t_2-i\tilde\epsilon)}
    {\sinh^2\left(-\pi F_{x+1}^+(-t_1+i\tilde\epsilon)+\pi F_{x+1}^-(-t_2-i\tilde\epsilon) +i\pi\right) } \r)^{1/2},
}
where $\tilde\epsilon = b/2 - \epsilon$. Using $\sinh(z+i\pi) = \sinh(-z)$ and further changing the integration variables $t_{1,2}\to -t_{1,2}$, we find
\es{}{
    \frac{i\tilde I^\epsilon_{+-}}{gN\pi^2} 
    &=
    \int dt_1 dt_2
    \l(  \frac{ F_x'^+(t_1+i\tilde\epsilon) F_x'^-(t_2-i\tilde\epsilon)}
    {\sinh^2\left(\pi F_x^+(t_1+i\tilde\epsilon)-\pi F_x^-(t_2-i\tilde\epsilon) \right) } \r)^{1/2} \\
    &\hspace{6em}
    \l(  \frac{ F_{x+1}'^+(t_1+i\tilde\epsilon) F_{x+1}'^-(t_2-i\tilde\epsilon)}
    {\sinh^2\left(\pi F_{x+1}^+(t_1+i\tilde\epsilon)-\pi F_{x+1}^-(t_2-i\tilde\epsilon)\right) } \r)^{1/2} \\
    &= \frac{iI^{\tilde\epsilon}_{+-}}{gN\pi^2} .
}
This is the same expression as \cref{eq:interaction-app}, simply with $\epsilon$ changed to $\tilde\epsilon$. The same argument follows for $I^{\varepsilon}_{-+}$. We can also similarly show this to be true for the diagonal terms $I^{\varepsilon}_{++}$ and $I^{\varepsilon}_{+-}$, where both fields live on the same Lorentzian fold.

As we have argued, the numerical value of the integral does not depend whether we choose the regulator to be $\epsilon$ or $\tilde\epsilon$, provided that both are small. This means that 
\begin{align}
    F^\pm(t) \to -F^\pm (-t \pm ib/2) + \frac{i}{2},
\end{align}
is a symmetry of action for any $2\epsilon< b \ll \beta_0$. 
This is effectively the dynamical KMS symmetry~\eqref{simpleshifts} for infinitesimal inverse temperature $b$. In the limit $\epsilon\to 0$, we can smoothly take $b\to 0$, recovering the KMS shift symmetry \eqref{eq:KMS-cont-together}. 

Naively, it may appear that the argument above also goes through for $\theta^\pm = 0$, resulting in the plausible symmetry $F^\pm(t) \to -F^\pm (-t \pm ib/2)$ that becomes the naive time reversal transformation in the limit $b\to 0$. However, this is clearly incorrect: we can explicitly check that our SK EFT in \eqref{Leff2} is not invariant under the naive time-reversal transformation.

The problem here is subtle. The fields $F^\pm(t,x)$ in the original Lorentzian effective action take real values. This means that the argument of $\sinh$ in \eqref{eq:interaction-app} has a small positive imaginary part $2i\pi\epsilon F'(t,x)$ for small $\epsilon$. However, taking $\theta^\pm=\theta$ and $b^\pm = b$, the transformation \eqref{eq:interaction-app} maps the argument of $\sinh$ to take an imaginary part $i\pi\le(\theta + (2\epsilon - b) F'^\pm(-t,x)\ri)$. Since $b > 2\epsilon$ and $F'^\pm(t,x) > 0$, the imaginary part lies slightly below $i\pi\theta$. If $\theta=1$, we can constantly deform the integration contour to bring the argument back to $2i\pi\epsilon F'(t,x)$ without encountering any singularities. However, for $\theta=0$, we would need to cross the singularity at the real line, meaning that the transformation $F^\pm(t) \to -F^\pm (-t \pm ib/2)$ maps the fields to a physically different sector than the original effective action and hence is not a symmetry of the theory.

\section{Proof of the second law}
\label{app:secondlaw}

In this appendix, we provide a proof of how the local second law of thermodynamics~\eqref{eq:second-law} emerges from the SK EFT framework, based on the unitarity constraints~\eqref{eq:unitarity} and the discrete symmetry~\eqref{eq:KMS-cont-together} combining the KMS symmetry with a discrete difference time translation. The discussion here is broadly based on~\cite{Glorioso:2016gsa, Armas:2024iuy}.

\subsection{Derivation of the entropy production rate}

Let us express the SK Lagrangian as
\begin{align}\label{eq:general-Lagrangian}
    \cL_\SK(F,Q)
    = \sum_{m=1}^{2R} (-i)^{m+1}\hat\cL_{(m)}Q^m,
\end{align}
where $\hat\cL_{(m)}$ are $F$-dependent symmetric multilinear differential operators acting on $Q^m$.\footnote{For example, $\hat\cL_{(2)}AB = \hat{\cal D}_1 A\, \hat{\cal D}_2 B + \hat{\cal D}_2 A\, \hat{\cal D}_1 B$ for some linear differential operators $\hat{\cal D}_{1,2}$ and arbitrary functions $A$ and $B$.} We will find it useful to split the operators into their time reversal even and odd parts, denoted $\hat\cL_{(m)}^e$ and $\hat\cL_{(m)}^o$ respectively.
Note that the expansion lacks a $ O(Q^0)$ term due to the first condition in~\eqref{eq:unitarity}. For clarity, we have chosen to truncate the expansion at $ O(Q^{2R})$, where $2R-1$ denotes the maximal odd power of $Q$ appearing in the Lagrangian working at a given order in derivatives. We can always take $R\to\infty$ in the following discussion.

The symmetry~\eqref{eq:KMS-cont-together} can be expressed in terms of the Lagrangian as
\begin{align}\label{eq:KMS-TT}
    \cL_\SK\big(\Theta F,\Theta Q\big) \Big|_{t\to -t}
    &= \cL_\SK(F,Q+i)\,.
\end{align}
Taking the imaginary part of this equation, we find
\es{eq:start-proof}{
    \Im\Big[\cL_\SK(\Theta F,\Theta Q)\Big]_{-t\to t}
    &= \frac{1}{2i}
    \Big[
    \cL_\SK\big(F,Q+ i\big)
    - \cL^*_\SK\big(F,Q- i\big)
    \Big] \\
    &= \frac{1}{2i}
    \Big[
    \cL_\SK\big(F,Q+i\big)
    + \cL_\SK\big(F,-Q+i\big)
    \Big]\,,
}
where in the second step we used the second unitarity condition in~\eqref{eq:unitarity}. We can now Taylor expand the two terms in~\eqref{eq:start-proof} in their noise arguments $\pm Q+ i$. Recall from~\eqref{eq:L1-expansion} that the $ O(Q^1)$ term in the expansion can be expressed in terms of the equations of motion and the entropy current operator
\begin{align}\label{eq:EC}
    - \lambda\hat\cL_{(1)}Q = Q\,(\text{EoM}) 
    + \dow_\mu\big( \hat\cS^\mu Q \big)\,,
\end{align}
where $\hat\cS^\mu$ is understood as a linear differential operator acting on $Q$. It follows that the first term in the expansion of~\eqref{eq:start-proof} takes the form
\begin{align}
    \frac{1}{2i} \Big[ {-}\hat\cL_{(1)}(Q+i) - \hat\cL_{(1)}(-Q+i)  \Big]
    = - \hat\cL_{(1)}1
    = \frac{1}{\lambda}\text{EoM}
    + \frac{1}{\lambda} \dow_\mu \cS^\mu,
\end{align}
where $\cS^\mu = \hat\cS^\mu 1$.
At next order, we find
\begin{align}
    \frac{1}{2i} \Big[ i\hat\cL_{(2)}(Q+i)^2 + i\hat\cL_{(2)}(-Q+i)^2  \Big]
    &= \hat\cL_{(2)}\left( Q^2 - 1^2\right)\,.
    \label{eq:second-order}
\end{align}
The complete expansion is given as
\begin{align}\label{eq:general-ec-eqn}
    \Im\Big[\cL_\SK(\Theta F,\Theta Q)\Big]_{-t\to t}
    = \frac{1}{\lambda}\text{EoM}
    + \frac{1}{\lambda}\p_\mu\cS^\mu
    - \sum_{m=2}^{2R}\hat\cL_{(m)}\Re(1+iQ)^m.
\end{align}

\subsection*{Gaussian noise}

Let us first analyse the simpler case of Gaussian noise, meaning that the Lagrangian is at most quadratic in noise fields. In this case $R=1$ and we have
\begin{align}
    \cL^{\text{Gaussian}}_\SK(F,Q) = -\hat\cL_{(1)}Q + i\hat\cL_{(2)}Q^2.
\end{align}
Imposing the symmetry~\eqref{eq:KMS-TT} on this Lagrangian implies that the operators must satisfy the constraints
\begin{align}\label{eq:gaussian-constraints}
    \hat\cL_{(1)}^e Q = - \hat\cL^e_{(2)}1 Q, \qquad 
    \hat\cL_{(1)}^o 1 = 0, \qquad 
    \hat\cL_{(2)}^o Q^2 = 0.
\end{align}
These conditions should be understood as being imposed for arbitrary $Q$.
The notation $\hat\cL_{(2)}1Q$ means that the bilinear operator $\hat\cL_{(2)}$ acts on 1 and $Q$. Correspondingly, the second condition means that $\hat\cL_{(1)}^o$ acting on 1 vanishes. Importantly, these constraints leave the even part of the quadratic operator $\smash{\hat\cL^e_{(2)}}$ completely arbitrary. However, to agree with the non-negativity unitarity constraint in~\eqref{eq:unitarity}, we must require
\begin{align}
    \hat\cL^e_{(2)}Q^2 \geq 0.
\end{align}
One may check that the our SK Lagrangian in~\eqref{Leff2} obeys these constraints. In fact, $\hat\cL^o_{(1)}$ and $\hat\cL^e_{(2)}$ precisely characterise the Schwarzian and dissipative parts of the action.

In the Gaussian case, we may evaluate~\eqref{eq:general-ec-eqn} at $Q\mapsto 1$, which kills the $\hat\cL^e_{(2)}$ contributions on the right-hand side, and we arrive at the desired on-shell statement of the second law of thermodynamics given in~\eqref{eq:second-law} with entropy production rate
\begin{align}
    \Delta 
    = \lambda \Im\Big[\cL^{\text{Gaussian}}_\SK(\Theta F,-1)\Big]_{-t\to t}
    = \lambda \Im\cL^{\text{Gaussian}}_\SK(F,1)
    = \lambda\cL_{(2)}1^2.
\end{align}
This agrees with the entropy production formula in~\eqref{eq:second-law} for theories with Gaussian noise. 

\subsection*{Non-Gaussian noise}

Let us now turn our attention to the general case. Plugging the Lagrangian~\eqref{eq:general-Lagrangian} into the symmetry constraint~\eqref{eq:KMS-TT}, we find
\begin{equation}
\begin{aligned}
    \sum_{m=1}^{2R} &\Big( \hat\cL_{(m)}^e
    - \hat\cL_{(m)}^o \Big)  (iQ)^m \\
    &=\sum_{m=1}^{2R} \Big( \hat\cL_{(m)}^e + \hat\cL_{(m)}^o\Big) (1-iQ)^m \\
    &=\sum_{m=1}^{2R}\sum_{r=0}^m \binom{m}{r}
    \Big( \hat\cL_{(m)}^e + \hat\cL_{(m)}^o\Big) 1^{m-r} (-iQ)^r \\
    &= \sum_{m=1}^{2R} \Big( \hat\cL_{(m)}^e + \hat\cL_{(m)}^o\Big) 1^m
    + \sum_{r=1}^{2R}\sum_{m=r}^{2R} \binom{m}{r}
    \Big( \hat\cL_{(m)}^e + \hat\cL_{(m)}^o\Big)1^{m-r} (-iQ)^r.
\end{aligned}
\end{equation}
Comparing the left-hand and right-hand sides for different powers of $Q$ and time reversal odd/even sectors, we find the constraints
\begin{equation}\label{eq:general-constraints}
\begin{alignedat}{2}
    \sum_{m=1}^{2R} \hat\cL_{(m)} 1^m &= 0, \qquad &
    \sum_{r=m+1}^{2R} \binom{r}{m} \hat\cL_{(r)} 1^{r-m} Q^m
    &= - \Big( 1 - \eta (-1)^m \Big) \hat\cL_{(m)} Q^m, 
\end{alignedat}
\end{equation}
which are valid for arbitrary $Q$. Here $\eta$ is $+1$ in the even sector and $-1$ in the odd sector.
One may check that they reduce to~\eqref{eq:gaussian-constraints} for $R=1$. These constraints can be solved for arbitrary $R$. Having done that, we can verify the identities (using Mathematica)
\begin{equation}\label{eq:summation-formulas}
\begin{aligned}
    \sum_{m=2}^{2R} \hat\cL^e_{(m)} \Re (1+i Q)^m
    &= - \sum_{k=1}^{R} (-1)^k \hat\cL^e_{(2 k)} \left(c^e_{(k)}-Q^{2 k}\right)\,, \\
    \sum_{m=2}^{2R} \hat\cL^o_{(m)} \Re (1+i Q)^m
    &= - \sum _{k=1}^{R-1} (-1)^k \hat\cL^o_{(2 k)} \left(c^o_{(k)}+Q^{2 k}\right)\,,
\end{aligned}
\end{equation}
where 
\begin{equation}\label{eq:coefficients}
\begin{aligned}
    c^e_{(k)} 
    &= \int_{-\infty }^{\infty }d q\,q^{2 k} \frac{\pi  \cosh (\pi  q)}{\sinh ^2(\pi  q)}
    = 2 (-1)^k \left(1-4^k\right) B_{2 k}\,, \\
    c^o_{(k)}
    &= \int_{-\infty }^{\infty }d q\,q^{2 k} \frac{\pi}{\sinh ^2(\pi  q)}
    = 2 (-1)^{k+1} B_{2 k}\,,
\end{aligned}
\end{equation}
where $B_n$ are the Bernoulli numbers.
Note that the integrals above are only well-defined for $k> 0$.
Let us define $c^e_{(0)} \equiv 1$ and $c^o_{(0)} \equiv -1$. Using these conventions, it follows that 
\es{}{
    \sum_{k=0}^\infty \frac{c^e_{(k)}}{(2k)!} \p^{2k}_Q \Big|_{Q\to 0}
    \lB \sum _{m=2}^{N} \hat\cL^e_{(m)} \Re (1+i Q)^m \rB
    &= 0, \\
    \sum_{k=0}^\infty \frac{-c^o_{(k)}}{(2k)!} \p^{2k}_Q \Big|_{Q\to 0}
    \lB \sum _{m=2}^{N} \hat\cL^o_{(m)} \Re (1+i Q)^m \rB
    &= 0.
}
The above operation essentially replaces $Q^{2k}$ with $c^e_{(k)}$ and $-c^o_{(k)}$ respectively, thereby setting the right-hand side of~\eqref{eq:summation-formulas} to zero. Note that, as defined,
\es{}{
    \sum_{k=0}^\infty \frac{c^e_{(k)}}{(2k)!} \p^{2k}_Q \Big|_{Q\to 0}
    {\cal H}
    &= {\cal H}, \qquad
    \sum_{k=0}^\infty \frac{-c^o_{(k)}}{(2k)!} \p^{2k}_Q \Big|_{Q\to 0}
    {\cal H}
    = {\cal H},
}
for any expression ${\cal H}$ that does not depend on $Q$. Therefore, using this on~\eqref{eq:general-ec-eqn}, we find
\es{}{
    &\sum_{k=0}^\infty \frac{c^e_{(k)}}{(2k)!} \p^{2k}_Q \Big|_{Q\to 0}\Im\!\Big[\cL^e_\SK(F,Q)\Big] 
    + \sum_{k=0}^\infty \frac{c^o_{(k)}}{(2k)!} \p^{2k}_Q \Big|_{Q\to 0}\Im\!\Big[\cL^o_\SK(F,Q)\Big]  \\
    &\qquad 
    = \frac{1}{\lambda}\text{EoM}
    + \frac{1}{\lambda}\p_\mu\cS^\mu.
}
Here we have used the fact that $\cL^e_\SK$ has eigenvalue $+1$ under time reversal, while $\cL^o_\SK$ has eigenvalue $-1$.
Lastly, the Lagrangian has no terms that are $ O(Q^0)$, therefore the summations above start from $k=1$ and we can reliably replace them with their integral representations in~\eqref{eq:coefficients}. This precisely reproduces the formula in~\eqref{eq:second-law-statement} with
\begin{align}\label{eq:general-Delta}
    \Delta = \lambda\int_{-\infty}^\infty dq
    \frac{\pi}{\sinh^2(\pi q)}
    \bigg( \cosh(\pi q)\Im\cL^e_\SK(F,q)
    + \Im\cL^o_\SK(F,q)
    \bigg)\,.
\end{align}
 Interestingly, in our theory, $\Im\cL^o_\SK$ is identically zero to all orders in $\lambda$, which is why we have used the simpler expression in~\eqref{eq:simple-Delta} in the main text.

\subsection{Local positivity of entropy production}

From the SK unitarity conditions~\eqref{eq:unitarity}, we know that $\Im I_\SK[F,Q]\geq 0$ for all field configurations. However, since the imaginary part $I_\SK^o$ flips sign for time reversed configurations, we must require that 
\begin{align}\label{eq:even-odd-positivity}
    \Im I^e_\SK[F,Q] \geq \big|\Im I^o_\SK[F,Q]\big|.
\end{align}
This condition guarantees that the total entropy production, integrated over space and time, is always non-negative. To wit,
\begin{align}
    S_{\text{tot}} = \int dtdx\, \Delta 
    = \lambda\int_{-\infty}^\infty dq\,
    \frac{\pi}{\sinh^2(\pi q)}
    \bigg( \cosh(\pi q)\Im I^e_\SK[F,q]
    + \Im I^o_\SK[F,q]
    \bigg)\,,
\end{align}
Since $\cosh(\pi q)\geq 1$ for all $q$,~\eqref{eq:even-odd-positivity} implies that $S_{\text{tot}}\geq 0$. This is essentially the generalisation of the argument presented near~\eqref{eq:total-entropy}.

However, hydrodynamics is built around the notion of local thermodynamic equilibrium. Therefore, we expect a stronger local version of the second law to hold in hydrodynamics: that the on-shell divergence of the entropy current $\Delta$ is non-negative at every point in spacetime. To achieve this, we note that 
\begin{align}
    \Im I_\SK = \int dtdx \Im\cL_\SK \geq 0.
\end{align}
Provided that $\Im\cL_\SK$ is smooth and falls of sufficiently fast at infinity, it follows that there must exist a smooth vector field  $V^\mu$ such that 
\begin{align}
    \Im\cL_\SK + \dow_\mu V^\mu \geq 0.
\end{align}
In other words, we can modify the Lagrangian with total derivative terms that leave the  SK action invariant, i.e.
\begin{align}
    \tilde\cL_\SK = \cL_\SK + i \dow_\mu V^\mu + \dow_\mu U^\mu,
\end{align}
such that
\begin{align}\label{eq:even-odd-positivity-L}
    \Im\tilde\cL_\SK(F,Q)\geq 0 
    \quad\Longleftrightarrow\quad 
    \Im\tilde\cL^e_\SK(F,Q)\geq \big|\Im \tilde\cL^o_\SK(F,Q)\big|.
\end{align}
The vector $U^\mu$ here is picked in such a way that the new Lagrangian continues to satisfy the symmetry constraint~\eqref{eq:KMS-TT}. We have explicitly checked the existence of such a $V^\mu$ and $U^\mu$ for our theory in the supplementary Mathematica notebook to $O(\lam^4)$.

We can now run the construction above to find the new entropy current $\tilde\cS^\mu$ using~\eqref{eq:L1-expansion} such that associated the entropy production rate $\tilde\Delta$ given by~\eqref{eq:general-Delta} is locally non-negative. One simply finds
\es{}{
    \tilde\cS^\mu 
    &= \cS^\mu 
    + \lambda\int_{-\infty}^\infty dq\,
    \frac{\pi}{\sinh^2(\pi q)}
    \bigg( \cosh(\pi q) V^{\mu e}(F,q) + V^{\mu o}(F,q)
    \bigg)\,.
}
Here we have introduced a splitting $V^{\mu} = V^{\mu e}+V^{\mu o}$, such that $\dow_\mu V^{\mu e}$ and $\dow_\mu V^{\mu o}$ are even and odd under time reversal respectively. Note that this means $V^{t e}$, $V^{x o}$ are odd under time reversal, while $V^{to}$, $V^{xe}$ are even. We have
\begin{align}
    \tilde\Delta = \lambda\int_{-\infty}^\infty dq\,
    \frac{\pi}{\sinh^2(\pi q)}
    \bigg( \cosh(\pi q)\Im\tilde\cL^e_\SK(F,q)
    + \Im\tilde\cL^o_\SK(F,q)
    \bigg)\,,
\end{align}
which is locally non-negative due to~\eqref{eq:even-odd-positivity-L}.

\subsection*{Constructing $U^\mu$}

To complete the proof of local entropy production, we only need to convince ourselves that a $U^\mu$ exists for any choice of $V^\mu$ ensuring~\eqref{eq:KMS-TT}. We implement the expansion
\es{eq:V-U-operators}{
    V^\mu = 
    \sum_{m=1}^{R} (-1)^{m+1}\hat\cV^\mu_{(m)}Q^{2m}, \qquad 
    U^\mu = 
    \sum_{m=1}^{R} (-1)^{m+1}\hat\cU^\mu_{(m)} Q^{2m-1}\,.
}
As before, we have introduced a truncation of the expansion parametrised by $R$. We will define it precisely below.
We introduce the even-odd splitting of operators, $\hat\cV^\mu_{(m)} = \hat\cV^{e\mu}_{(m)}+\hat\cV^{o\mu}_{(m)}$, such that $\dow_\mu\hat\cV^{e\mu}_{(m)}$ and $\dow_\mu\hat\cV^{o\mu}_{(m)}$ are even and odd under time reversal respectively. We introduce a similar splitting for the operators $\hat\cU^\mu_{(m)}$.

Plugging the decomposition~\eqref{eq:V-U-operators} into the symmetry condition~\eqref{eq:KMS-TT}, we find the constraints akin to~\eqref{eq:general-constraints}, i.e.
\begin{equation}\label{eq:general-constraints-V}
\begin{aligned}
    \sum_{m=1}^{R} \hat\cU_{(m)} 1^{2m-1}
    &= - \sum_{m=1}^{R} \hat\cV_{(m)} 1^{2m}, \\
    (1 + \eta) \hat\cU_{(m)} Q^{2m-1}
    &= 
    - \sum_{r=m}^{R} \binom{2r}{2m-1} \hat\cV_{(r)} 1^{2r-2m+1} Q^{2m-1}
    - \sum_{r=m+1}^{R} \binom{2r-1}{2m-1} \hat\cU_{(r)} 1^{2r-2m} Q^{2m-1}, \\
    (1 - \eta) \hat\cV_{(m)} Q^{2m}
    &= - \sum_{r=m+1}^{R} \binom{2r-1}{2m} \hat\cU_{(r)} 1^{2r-2m-1} Q^{2m}
    - \sum_{r=m+1}^{R} \binom{2r}{2m} \hat\cV_{(r)} 1^{2r-2m} Q^{2m},
\end{aligned}
\end{equation}
where $\eta=+1$ in the even sector and $\eta=-1$ in the odd sector. We have dropped the spacetime label $\mu$ for clarity.
We have to solve these constraints to find $\hat\cU_{(m)}$ operators in terms of $\hat\cV_{(m)}$ operators. In particular, substituting $m\mapsto R$ in the last condition in~\eqref{eq:general-constraints-V} for $\eta = -1$, we find a consistency condition on the $\hat\cV_{(R)}$ operator, i.e.
\begin{align}
    \hat\cV^{o}_{(R)}Q^{2R} = 0.
\end{align}
In other words, we can define $R$ as the lowest number for which the operator $\hat\cV^{o}_{(R)}$ vanishes. In other words, $R$ is the lowest number for which both the time reversal even part of $\hat\cV^{t}_{(R)}$ and the time reversal odd part of $\hat\cV^{x}_{(R)}$ vanishes.

In the trivial $R=1$ case, we find the constraints
\begin{equation}\label{eq:R1cons}
\begin{aligned}
    \hat\cU^{o}_{(1)} 1
    &= 0, \qquad
    \hat\cU_{(1)}^{e} Q
    = - \hat\cV_{(1)}^{e} 1 Q, \qquad 
    \hat\cV^{o}_{(1)}Q^2 = 0,
\end{aligned}
\end{equation}
which admit a solution
\begin{equation}\label{eq:R1soln}
    \hat\cU_{(1)} Q = - \hat\cV_{(1)}^{e} 1 Q.
\end{equation}
Note that this is not the most general solution of the constraints~\eqref{eq:R1cons}. However, we only need to illustrate the existence of some $\hat\cU^\mu_{(m)}$ operators that solve the constraints and not necessarily find the most general ones. The constraints for $R=2$ case are more interesting:
\es{}{
    \hat\cU_{(1)}^{e} Q
    &= 
    - \hat\cV_{(1)}^{e} 1 Q
    + \hat\cV_{(2)}^{e} 1^{3} Q, \\
    \hat\cU_{(2)}^{e} Q^{3}
    &= 
    - 2\hat\cV_{(2)}^{e} 1 Q^{3}, \\
    \hat\cU_{(2)}^{o} 1 Q^{2}
    &= - \frac{2}{3}\hat\cV_{(1)}^{o} Q^{2}, \\
    \hat\cV^{o}_{(2)}Q^4 &= 0,
}
which has the solution
\es{}{
    \hat\cU_{(1)} Q
    &= 
    - \hat\cV_{(1)}^{e} 1 Q
    + \hat\cV_{(2)}^{e} 1^{3} Q
    - \frac{1}{3} \hat\cV_{(1)}^{o} 1 Q, \\
    \hat\cU_{(2)} Q^3 
    &= - 2\hat\cV_{(2)}^{e} 1 Q^{3}
    - \frac{2}{3}\Big( 3Q \hat\cV_{(1)}^{o} Q^{2} - 3Q^2\hat\cV_{(1)}^{o} 1Q + Q^3\hat\cV_{(1)}^{o}1^{2} \Big)\,.
}
We have checked in the Mathematica file that this procedure to generate $U^\mu$ can be extended to arbitrarily large values of $R$. It seems likely that there exists an explicit proof by induction, but we leave this exercise for future work.

\section{Collection of formulas}

\subsection{SK EFT Lagrangian at higher order in the gradient expansion}
\label{app:highorder}

Here we record various formulas to higher order in the gradient expansion. The formulas are also available to even higher order in a Mathematica file attached to the \texttt{arXiv} submission and also in the GitHub repository~\cite{github_syk_hydro_2026}. The SK effective Lagrangian up to $O(\lam^2)$ in the discrete form is given as
\es{Ldisc}{
\frac{{L}_\SK}{\cN}
&=\sum_x\Bigg[ \frac{F' Q'}{D}+\frac{i \left(R F'-Q
   G'\right) \left(R F'-(Q-i) G'\right)}{F'} \\
   &\qquad
   +\lambda ^2 \left(\frac{G' R'}{4 D}+\frac{i \left(-42 \pi ^2 Q
   (Q-i)+7 \pi ^2+15\right) R^2 G'^2}{60 F'}\ri.\\
   &\qquad\qquad
   +\frac{(1+2 i Q) \left(-15+14 \pi
   ^2 Q (Q-i)\right) R G'^3}{60 F'^2}  \\
   &\qquad\qquad 
   -\frac{7}{60} i \pi ^2 R^4 F'
   -\frac{i Q (Q-i)\left(-15+7 \pi ^2 Q (Q-i)\right) G'^4}{60F'^3} \\
   &\qquad\qquad 
   + \le.\frac{7}{30} \pi ^2 (1+2 i Q) R^3 G'
   + \dow_{tt}\left(\frac{3(1+iQ) Q)}{4 \pi ^2F'}
   \right)
   \right)+O(\lam^4)\Bigg]\,,
}
where $\cN = 2\pi^3 gN/3$. The same results in the continuum form are given as
\es{}{
\frac{{\cal L}_\SK}{\cN/\lambda} 
&=
\frac{\partial_t F\,\partial_t Q}{D}
+\frac{i\left(\partial_t F\,\partial_x Q-\partial_{tx}F\,Q\right)\left(\partial_t F\,\partial_x Q-(Q-i)\partial_{tx}F\right)}{\partial_t F}\\
&\qquad
+\lambda ^2\Bigg[
-\frac{7i\pi^2}{60}\partial_t F\,(\partial_x Q)^4 
+\frac{7\pi^2}{30}\partial_{tx}F\,(1+2iQ)\,(\partial_x Q)^3
\\
&\qquad\qquad
+(\partial_x Q)^2\left(
\frac{i(\partial_{tx}F)^2}{60\,\partial_t F}
\Big(7\pi^2+15 -42\pi^2 Q(Q-i)\Big)
+\frac{i}{8}\,\partial_{txx}F
\right)\\
&\qquad\qquad 
+ \frac{i(2Q-i)\,\partial_x Q}{120}
\left(\frac{(\partial_{tx}F)^3}{(\partial_t F)^2}\left(-30+28\pi^2 Q(Q-i)\right)
-5\partial_{txxx}F\right)
\\
&\qquad\qquad 
+ \partial_{xxx}Q\left(
\frac{i\partial_x Q}{12}\partial_t F
- \frac{1+2iQ}{24}\,\partial_{tx}F
\right)  \\
&\qquad\qquad 
+\partial_{xx}Q\left(
\frac{(\partial_{tx}F)^2(1+2iQ)}{8\,\partial_t F}
-\frac{i}{4}\partial_{tx}F\,\partial_x Q
\right) \\
&\qquad\qquad 
-\frac{i\,Q(Q-i)}{60(\partial_t F)^3}
(\partial_{tx}F)^4\Big(7\pi^2 Q(Q-i)-15\Big)  \\
&\qquad\qquad 
-\frac{i\,Q(Q-i)}{24(\partial_t F)^2}
\partial_{tx}F\Big(
3\partial_{txx}F\,\partial_{tx}F
-2\partial_t F\,\partial_{txxx}F
\Big)  \\
&\qquad\qquad
+ \p_{xx}\left(\frac{
\partial_t F\,\partial_{t}Q
}{8D} \right)
+ \p_{tt}\left(\frac{3 (1+iQ)Q}{4 \pi ^2\p_t F} \right)
\Bigg]
   +O(\lam^4)\,.
}
In the expressions above, we have isolated some total derivative terms that do not contribution to the effective action. These are an artefact of the prescription to go from discrete to continuum variables that we have explained in the main text. 

After performing several total derivative improvements, the Lagrangian takes an improved form where its imaginary part is manifestly non-negative
\es{eq:improved-L-answer}{
\frac{\tilde\cL_\SK}{\cN/\lambda}
&= \frac{\partial_t F\,\partial_t Q}{D}
-\partial_{tx}F\,\partial_x Q
+\frac{(\partial_{tx}F)^2 Q}{\partial_t F} \\
&\quad
+ \frac{\lambda^2}{120}\Bigg[
\frac{5\partial_{tx}F}{\partial_t F}\Big(
Q\partial_{txxx}F
- 3\partial_x Q \partial_{txx}F
+ 3\partial_{xx}Q\partial_{tx}F
- \partial_{xxx}Q\partial_{t}F
\Big) \\
&\qquad\qquad
+ 5\left(\frac{6(\partial_{tx}F)^3}{(\partial_t F)^3}
-\frac{3\partial_{txx}F\partial_{tx}F}{(\partial_t F)^2}
+\frac{\partial_{txxx}F}{\partial_t F}\right)
\Big( Q\partial_{tx}F - \dow_tF\dow_xQ\Big) \\
&\qquad\qquad
- \frac{28\pi^2}{(\dow_tF)^3}\partial_{tx}F 
\Big(Q\partial_{tx}F-\dow_tF\partial_x Q\Big)^3
\Bigg] + O(\lam^4) \\
&+ \frac{i}{\dow_tF}\Bigg\{
Q\partial_{tx}F - \partial_t F\,\partial_x Q \\
&\quad
+\frac{\lambda^2}{240}
\Bigg[
10\Big(Q\partial_{txxx}F
- 3\partial_x Q \partial_{txx}F
+ 3\partial_{xx}Q\partial_{tx}F
-\partial_{xxx}Q \partial_t F \Big)
\\
&\qquad\qquad
+ \left(\frac{30(\partial_{tx}F)^2}{(\dow_tF)^2} 
- \frac{15\partial_{txx}F}{\dow_tF} \right)
\Big(Q\partial_{tx}F- \dow_tF\partial_x Q \Big)
\\
&\qquad\qquad
- \frac{14\pi^2}{(\partial_t F)^2} 
\Big( Q\partial_{tx}F - \dow_tF\partial_x Q\Big)
\left(\Big(Q\partial_{tx}F-\dow_tF\partial_x Q\Big)^2
- (\partial_{tx}F)^2 \right)\Bigg]  + O(\lam^4) \Bigg\}^2.
}
We use this form to read off the improved entropy current whose divergence is manifestly non-negative using~\eqref{eq:simple-Delta}.

\subsection{Noether currents and the entropy current} \label{app:quantumNoether}

The time components of the energy currents was given in~\eqref{EtSch}.
The space components of the average and difference energy currents are 
\es{Exhigher}{
\frac{{\cal E}^x}{\cN}
&=
\partial_t F\Big(2i\,\partial_t F\,\partial_x Q - (1+2iQ)\,\partial_{tx}F\Big) \\
&\qquad 
+\lambda ^2 \Bigg[
-\frac{7i\pi^2}{15} (\dow_tF)^2 (\dow_xQ)^3
+\frac{7\pi^2}{10} \dow_tF\dow_{tx}F (1+2iQ) (\dow_xQ)^2
\\
&\qquad\qquad 
+ \frac{i}{30} \dow_xQ \bigg(
15\dow_tF \dow_{txx}F 
+ 7 \pi ^2 (\dow_{tx}F)^2 (1+i6 Q (1+iQ))\bigg) \\
&\qquad\qquad 
+ \frac{i}{6} (\dow_tF)^2 \dow_{xxx}Q
+ \frac{7 \pi ^2 (\dow_{tx}F)^3 (1+2 i Q) Q(Q-i)}{30\dow_tF}
\\
&\qquad\qquad 
- \frac{1}{12}
\left(3 \dow_{tx}F \dow_{txx}F + \dow_tF \dow_{txxx}F\right) 
(1+2iQ)
\Bigg]
+O(\lam^4)\,,\\[6pt]
\frac{{\cal E}_q^x}{\cN}
&= 4 i \dow_tF \dow_tQ\dow_xQ 
+ \Big( \dow_tF \dow_{tx}Q
- 3\dow_{tx}F \dow_tQ \Big) (1+2 i Q) \\
&\qquad 
+ 2 \left( \frac{\dow_{tx}F\dow_{tt}F}{\dow_t F}- \dow_{ttx}F\right) 
   (1+i Q) Q+O(\lam^2)\,.
}
The time component of the stress tensor agrees with ${\cal E}^\mu_q$, the space component is given~by
\es{taux}{
\frac{\tau^t_{\,\,x}}{\cN}
&= \frac{\dow_xF\dow_tQ+\dow_tF\dow_xQ}{D}
- \dow_xQ \left( \half\dow_{xx}F
   +\frac{\dow_xF \dow_{tx}F}{\dow_tF} \right) (1+2 iQ) 
   + 2 i \dow_xF (\dow_xQ)^2\\
&\qquad
+ \frac{1}{2} \dow_xF \dow_{xx}Q (1+2 i Q) 
+ \frac{\dow_{xx}F \dow_{tx}F - \dow_xF\dow_{txx}F}{\dow_tF} (1+i Q) Q
+ O(\lam^2)\,,\\[6pt]
\frac{\tau^x_{\,\,x}}{\cN}
&= -\frac{\dow_tF\dow_tQ}{D}
- \dow_tQ 
\frac{\dow_xF \dow_{tx}F}{\dow_tF} (1+2 iQ)
+ i \dow_tF (\dow_xQ)^2\\
&\qquad
+ \dow_xQ \left(i\dow_xF \dow_tQ
- \frac{1}{2} \dow_{tx}F (1+2 i Q)\right)
+ \frac{1}{2} \dow_xF (1+2 i Q) \dow_{tx}Q \\
&\qquad 
+\frac{\dow_{tx}F \dow_{tt}F- \dow_tF \dow_{ttx}F}{(\dow_tF)^2}
   \dow_xF (1+i Q) Q 
   + O(\lam^2)\,.
}

The $\SL(2,\mathbb{R})$ shift current was given in schematic form in~\eqref{SL2schematic}, its explicit form is:
\es{SL2concrete}{
\frac{{\cal G}^t}{\cN}
&=
\p_x\delta_{\boldsymbol 0}(x)
\left(\frac{\dow_{tx}F}{\dow_tF} (1+i Q) Q
- \frac{1}{2} (1+2 iQ) \dow_xQ\right) \\
   &\qquad
   +\delta_{\boldsymbol 0}(x) \Bigg[
   \frac{\dow_tQ}{D}
   - \frac{\dow_{txx}F (1+iQ)Q}{\dow_tF} 
   + 2i (\dow_xQ)^2 \\
   &\qquad\qquad\qquad\qquad
   + \left( \frac{1}{2} \dow_{xx}Q
   - \frac{\dow_{tx}F}{\dow_tF}\dow_xQ \right) (1+2 i Q)
   \Bigg]
   +O(\lam^2)\,,\\[6pt]
   \frac{{\cal G}^x}{\cN}
&=\delta_{\boldsymbol 0}(x) \Bigg[
   \frac{\dow_{tx}F \dow_{tt}F -\dow_tF\dow_{ttx}F}{(\dow_tF)^2} (1+i Q) Q 
   + i\dow_tQ\dow_xQ \\
   &\qquad\qquad
   + \left( \frac{1}{2} \dow_{tx}Q
   -\frac{\dow_{tx}F \dow_tQ}{\dow_tF} \right) (1+2 i Q)  \Bigg]+O(\lam^2)\,.
}

The new entropy current with positive semidefinite local divergence is given by
\es{}{
\frac{\tilde\Sc^\mu}{\cN} &=\begin{pmatrix}
\displaystyle \frac1D \dow_t F \\[4pt]
\displaystyle - \partial_t\partial_x F 
- \left(\frac{(\partial_{tx} F)^3}{8(\partial_t F)^2} 
+ \frac{\partial_{tx} F\partial_{txx} F}{4\partial_t F} 
+ \frac{1}{12}\partial_{txxx} F\right)\lambda^2
\end{pmatrix}+O(\lam^4)\,,\\
\frac{\tilde\Delta}{\cN}&= 
\frac{(\partial_{tx} F)^2}{\partial_t F}
+ \frac{\lambda^2}{4}\frac{\partial_{tx} F}{\partial_t F}\left(
\frac{(\partial_{tx} F)^3}{(\partial_t F)^2} - 
\frac{\partial_{txx} F\,\partial_{tx} F}{2\partial_t F}
+ \frac{\partial_{txxx} F}{3} \right)
+ O(\lam^4)\,.
}
The result for entropy production rate can be expressed as a non-negative integral of a whole square:
\es{}{
\frac{\tilde\Delta}{\cN}
&= \int_{-\infty}^\infty dq \,\frac{\cosh(\pi q)}{\sinh^2(\pi q)}\frac{q^2}{\dow_tF} \\
&\qquad
\Bigg[
\partial_{tx}F
+\frac{\lambda^2}{8}
\Bigg(
\frac{(\partial_{tx}F)^3}{(\dow_tF)^2}\frac{15+7\pi^2(1-q^2)}{15}
- \frac{\partial_{tx}F\partial_{txx}F}{2\dow_tF} 
+ \frac{\partial_{txxx}F}{3}
\Bigg) + O(\lam^4) \Bigg]^2,
}
which is obtained after substituting the improved $\Im\tilde\cL_\SK$ from~\eqref{eq:improved-L-answer} into the formula~\eqref{eq:simple-Delta}.

\bibliographystyle{JHEP}

\bibliography{biblio}

\end{document}